\documentclass[11pt,a4paper]{article}
\pdfoutput=1

\usepackage{jcappub}
\usepackage{amsfonts}
\usepackage{amsmath}
\usepackage{amssymb}
\usepackage{aas_macros}
\usepackage{graphicx}
\usepackage{color}
\usepackage{float}
\usepackage{bm}
\usepackage{natbib}
\usepackage{hyperref}
\usepackage{longtable}

\def\lsim{~\rlap{$<$}{\lower 1.0ex\hbox{$\sim$}}}
\def\bsim{~\rlap{$>$}{\lower 1.0ex\hbox{$\sim$}}}

\def\apjl{Astrophys.\ J.\ Lett.}

\def\aap{Astron.\ Astrophys.}

\def\apjs{Astrophys.\ J. Supp.}

\newcommand{\be}{\begin{equation}}
\newcommand{\ee}{\end{equation}}

\let\ln\relax
\DeclareMathOperator{\ln}{ln}
\DeclareMathOperator{\tr}{tr}
\let\det\relax
\DeclareMathOperator{\det}{det}

\def\mathbi#1{\textbf{\em #1}}

\def\refeq#1{Eq.~(\ref{eq:#1})}

\def\ntvh{\tilde{\mathbi{n}}}

\def\kvh{\hat{\mathbi{k}}}
\def\nvh{\hat{\mathbi{n}}}

\def\nth{\tilde{\mathit{n}}}
\def\nh{\mathit{\hat{n}}}

\def\vk{\mathbi{k}}

\def\vv{\mathbi{v}}

\def\vx{\mathbi{x}}
\def\tvx{\tilde{\bm{x}}}

\def\grad{\mathbi{$\nabla$}}

\def\Sigtau{\Sigma_{\tau_o}}
\def\Sigtauz{\Sigma_{\tau_o;\tilde z}}

\def\omtau{\omega_{\tau_o}}
\def\tomtau{\tilde{\omega}_{\tau_o}}
\def\dgobs{\delta_g^{\rm obs}}

\def\veldiv{\theta}
\def\HF{H_0^\text{\tiny CFC}}

\def\DAF{d_A^\text{\tiny CFC}}
\def\at{a_t}

\def\intk{\int_{\bm k}}

\newcommand{\abs}[1]{\left\vert#1\right\vert}

\newcommand{\eps}{\varepsilon}

\def\emph#1{\textit{#1}}

\def\comment#1{}
\definecolor{RedWine}{rgb}{0.743,0,0}
\definecolor{RoyalBlue}{rgb}{0.25,.41,.88}
\definecolor{ForestGreen}{rgb}{.13,.54,.13}
\definecolor{DeepPurple}{rgb}{.72,.18,1}

\allowdisplaybreaks

\title{Covariant Decomposition of The Non-Linear Galaxy Number Counts and Their Monopole}

\author[a]{Yonadav Barry Ginat,}
\author[a,b]{Vincent Desjacques,}
\author[c]{Donghui Jeong,}
\author[d]{and Fabian Schmidt}

\affiliation[a]{Physics department, Technion, 3200003 Haifa, Israel}
\affiliation[b]{Asher Space Research Institute, Technion, 3200003 Haifa, Israel}
\affiliation[c]{Department of Astronomy and Astrophysics and Institute for Gravitation and the Cosmos, The Pennsylvania State University, University Park, PA 16802, USA}
\affiliation[d]{Max-Planck-Institut f\"ur Astrophysik, Karl-Schwarzschild-Stra\ss e~1, 85748 Garching, Germany}

\emailAdd{ginat@campus.technion.ac.il}
\emailAdd{dvince@physics.technion.ac.il}
\emailAdd{djeong@psu.edu}
\emailAdd{fabians@mpa-garching.mpg.de}

\date{\today}

\label{firstpage}

\abstract{We present a fully non-linear and relativistically covariant expression for the observed galaxy density contrast. Building on a null tetrad tailored to the cosmological observer's past light cone, we find a decomposition of the non-linear galaxy over-density into manifestly gauge-invariant quantities, each of which has a clear physical interpretation as a cosmological observable. This ensures that the monopole of the galaxy over-density field (the mean galaxy density as a function of observed redshift) is properly accounted for. We anticipate that this decomposition will be useful for future work on non-linearities in galaxy number counts, for example, deriving the relativistic expression for the galaxy bispectrum. We then specialise our results to conformal Newtonian gauge, with a Hubble parameter either defined globally or measured locally, illustrating the significance of the different contributions to the observed monopole of the galaxy density.
}

\begin{document}

\maketitle

\section{Introduction}
\label{sec:intro}

The observed clustering of galaxies provides a wealth of information on the constituents of the Universe, as well as on the nature of the processes that seeded the primordial fluctuations \cite{Laureijs:2011gra,Dore:2014cca,Aghamousa:2016zmz}. To exploit this information in galaxy clustering, however, one must understand how galaxies trace the underlying matter distribution. This relation includes not only the rest-frame galaxy bias, but also the selection and projection effects that appear in the mapping between the rest-frame and the observed galaxy distributions \cite[see][for a detailed review of all these contributions]{Desjacques:2016bnm}.
In linear theory, Refs.~\cite{Yoo:2009au,ChallinorLewis2011,Bonvin:2011bg,Jeong:2011as} have developed a relativistic formalism for incorporating the projection effects from gravitational lensing (leading to magnification of light and distortions of the galaxy positions), peculiar velocities, and wide angle effects
\cite{Sachs:1967er,Dyer:1973zz,dyer/roeder:1974,weinberg:1976,Kaiser:1987qv,Sasaki:1987ad,Futamase:1989hba,bartelmann/schneider:2001,Suto:1999id,Matsubara:1999du,Matsubara:2000pr,Bonvin:2005ps} onto the observed galaxy density.

In related studies, Refs.~\cite{SchmidtJeong2012,JeongSchmidt2014} introduced the cosmic ruler and cosmic clock observables, for situations where the intrinsic shape or the proper time are known at the source's rest frame. The deviation of the observed shape (or time) from the intrinsic shape (or the known time), therefore, forms a set of observables, denoted by ${\cal C}$, ${\cal B}_i$, ${\cal A}_{ij}$, respectively, for radial-radial, radial-tangential, and tangential distortion, and ${\cal T}$ for the shift in observed time. As observables, they are gauge-invariant and respect the equivalence principle by construction; that is, they are invariant under a constant or pure gradient shift of the gravitational potentials.

In this paper, we extend the formalism in the following two ways.
First, we apply the cosmic ruler framework of Refs.~\cite{SchmidtJeong2012,JeongSchmidt2014} to the galaxy number counts, and find a covariant decomposition for the
observed galaxy density contrast $\dgobs$ into an expression involving relatively simple, individually observable quantities, each of which has a clear physical meaning on its own. The new expression for the galaxy density contrast is not restricted to small, linear perturbations around an FLRW (Friedmann-Lema\^itre-Robertson-Walker) background, and is manifestly gauge-invariant at any order in perturbation theory. To do so, we start from a light-cone-based approach involving a null tetrad defined appropriately as in the Newman-Penrose formalism (see \cite{Penrose:1984uia,Newman:1961qr} and also \cite{ellis/etal:1985,stoeger/etal:1992,maartens/matravers:1994} for earlier applications of this formalism to the Large Scale Structure). The advantages of such a method lie in its intrinsically covariant nature, and in the manifestly apparent light cone structure of the observed galaxy distribution that emerges from it.

Second, it is important to take into account observer-related terms in $\dgobs$ in order to ensure its general covariance. As an example, the gravitational redshift depends on the potential difference of the source and observer, and thus requires the inclusion of the gravitational potential at the observer's position. Likewise, the epoch of observation is determined by a fixed \emph{proper} time of the observer, rather than fixed coordinate time.
These ``observer terms'' are the second main subject of our paper. Most of the early computations of projection effects did not account for all the observer terms, which are essential for the interpretation of the monopole (see e.g. \cite{Grimmetal2020,Desjacques:2020zue,CastorinaDiDio2021} for different views on this question). Below, we will take into account the observer terms properly and clarify their relation to the cosmic-ruler observables, and to the choice of background.

The paper is organised as follows. In Section \ref{sec:observed galaxy counts}, we present a derivation of the observed galaxy counts based on the Newman-Penrose formalism that illuminates the connection between different approaches in the literature. In Section \ref{sec:non-linear delta g}, we apply our results to derive a fully non-linear expression of the observed galaxy over-density in terms of cosmic clocks and rulers. In Section \ref{sec:cNresult}, we specialise our result to the conformal Newtonian gauge in linear theory and spell out the connection among the observer terms, the cosmic rulers and the choice of {\it background} Hubble parameter.
In Section \ref{sec:observed low multipoles}, we compute the low multipoles of the observed angular galaxy power spectrum and emphasise their dependence on the observer terms. We summarise our results in Section \ref{sec:discussion}. We adopt a signature $(-,+,+,+)$ throughout this paper, and in appendix \ref{appendix:symbol list} we give a list of the main symbols and notation used in this paper, for the reader's convenience.

\section{Observed galaxy counts}
\label{sec:observed galaxy counts}

Let $j = j^\mu \partial_\mu$ be the galaxy current 4-vector with components $j^\mu= n_g u_g^\mu$, where $n_g$ is the physical number density of galaxies in their rest frame and $u_g$ is the galaxies' 4-velocity.
In analogy with electromagnetism, the Hodge dual $\star j = i_j\,\omega$ (where $i_j$ is the interior product and $\omega$ is the volume form of the spacetime manifold $\mathbb{M}$) is the associated galaxy current 3-form.
The total number of galaxies $N(\Sigma)$ on a three-dimensional hyper-surface $\Sigma\subset \mathbb{M}$ (be it space-like or null) is therefore
\begin{equation}
    \label{eq:galaxy number count Hodge}
     N(\Sigma) = \int_{\Sigma} \,\star j = \int_\Sigma i_j\,\omega \;.
\end{equation}
When $\Sigma$ is identified with the three-dimensional past light cone, $\Sigtau$, of an observer $\mathcal{O}$ at given proper time $\tau_o$, $N(\Sigtau)$ gives the number of galaxy world-lines crossing $\Sigtau$, that is, the total number of galaxies seen by $\mathcal{O}$.

In Section \ref{sec:invariant volume forms}, we take a geometrical approach and express Eq.~\eqref{eq:galaxy number count Hodge} in a coordinate-independent way in terms of null vectors, elucidating a geometrical interpretation of $N(\Sigtau)$. We then derive an expression for $N(\Sigtau)$ as a function of the coordinates inferred by the observer in Section \ref{sec:volume form in inferred coordinates}.

\subsection{Invariant volume forms and geometrical interpretation}
\label{sec:invariant volume forms}

To proceed further, let $\omtau$ be the induced volume form on $\Sigtau$. The total number of galaxies in $\Sigtau$ is related to the number density as follows:
\begin{equation}\label{eq:galaxy number count}
  N(\Sigtau) = \int_{\Sigtau} n_{\rm g,lc}(x)~\omtau(x) \;,
\end{equation}
where $n_{\rm g,lc}(x)$ is an invariant number density on the light cone to be determined in terms of the current 4-vector $j$. Equation \eqref{eq:galaxy number count} is the same as equation \eqref{eq:galaxy number count Hodge} for $\Sigma=\Sigtau$, which can be seen by restricting the Hodge dual $\star j$ to $\Sigtau$, that is, by restricting the domain  of $\star j$ to vectors that are tangent to $\Sigtau$.

To compute $n_{\rm g,lc}(x)$ and $\omtau$, it is easier to introduce a null tetrad \cite{Newman:1961qr} tailored to $\Sigtau$, rather than to work in the usual comoving coordinates $x^\mu=(\eta,x^i)$. To this end, we define four orthonormal null vectors
$k = k^\mu\partial_\mu$, $l = l^\mu\partial_\mu$, $m = m^\mu\partial_\mu$, $\bar m = \bar m^\mu\partial_\mu$,
whose inner products vanish, except for
\begin{align}
\label{eq:nulltetrad}
\begin{aligned}
  & k\cdot l =  k^\mu l_\mu = -1 \\ &
  m\cdot \bar m = m^\mu\bar{m}_\mu = 1 \;.
\end{aligned}
\end{align}
The real-valued vectors $k$ and $l$ are future-null directed, whereas the complex-valued vectors $m$ and $\bar m$ describe the two types of circular polarisation.
In terms of these vectors, the full metric tensor is given by
\begin{equation}
  \mathbf{g} = -\mathrm{d}U\otimes\mathrm{d}V - \mathrm{d}V\otimes\mathrm{d}U+ \mathrm{d}\zeta\otimes\mathrm{d}\bar{\zeta} +\mathrm{d}\bar{\zeta}\otimes\mathrm{d}\zeta\;,
\end{equation}
where $\otimes$ denotes the usual tensor product. The 1-forms $\mathrm{d}V$ and $\mathrm{d}U$ are defined such that $\mathrm{d}V(k)\equiv 0$ and $\mathrm{d}U(l)\equiv 0$, that is, $k$ is tangent to hyper-surfaces of constant $V$, and likewise of $l$ and $U$. We further constrain $\mathrm{d}V$ and $\mathrm{d}U$ such that they are dual to $l$ and $k$, respectively: $\mathrm{d}V(l)= 1$ and $\mathrm{d}U(k)= 1$. As a result, we must have $\mathrm{d}V\equiv -k_\mu \mathrm{d}x^\mu$ and $\mathrm{d}U\equiv -l_\mu \mathrm{d}x^\mu$. Likewise, we require that the 1-forms $\mathrm{d}\zeta$ and $\mathrm{d}\bar\zeta$ be dual to $m$ and $\bar{m}$, respectively. In addition, we define the observer's past light cone $\Sigtau$ as some hyper-surface of constant $V$ in $\mathbb{M}$.
Note that our null coordinate $V$ is similar to that introduced in the geodesic-light-cone (GLC) coordinate system \cite[e.g.][]{Gasperini:2011us,Fanizza:2013doa,Fanizza:2015swa,Fleury:2016htl,Mitsouetal2020}.
However, the GLC coordinate basis vectors are not all null vectors and, therefore, do not satisfy the relations \eqref{eq:nulltetrad}.

\begin{figure}
    \centering
    \includegraphics[width=0.7\textwidth]{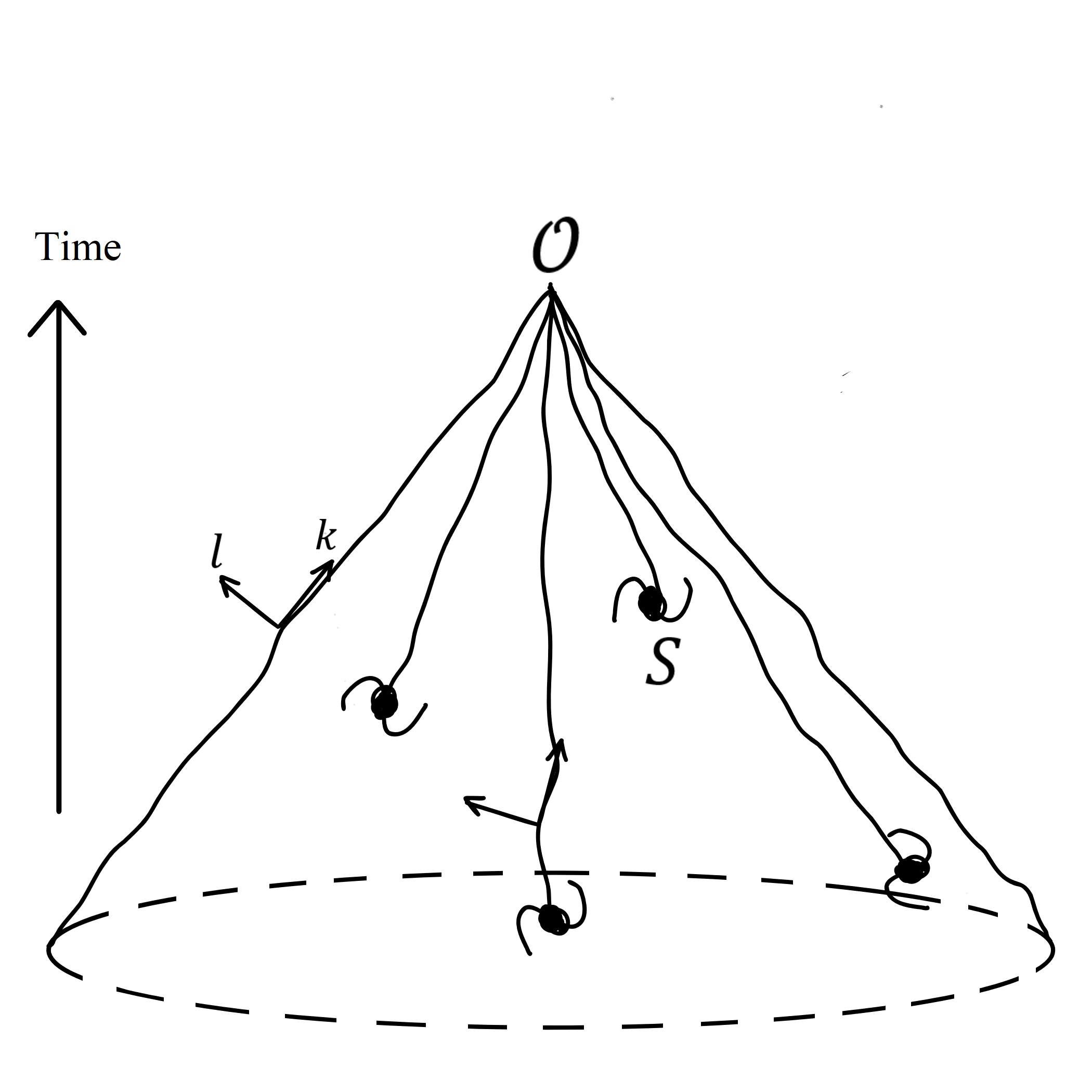}
    \caption{The (perturbed) past light cone of the observer $\mathcal{O}$ and the null tetrad.
      Null geodesics connecting the observer to the sources $\mathcal{S}$ (galaxies) are also shown. Illustration by Lea Reuveni.}
    \label{fig:cone}
\end{figure}

Let us now write the full volume form as
\begin{equation}
\label{eq:full volume form}
\omega=\mathrm{i}~\mathrm{d}U \wedge \mathrm{d}V\wedge \mathrm{d}\zeta \wedge\mathrm{d}\bar{\zeta} \;.
\end{equation}
Substituting this expression into Eq.~\eqref{eq:galaxy number count Hodge}, and using the properties of the tetrad vectors and exterior forms, yields an expression for the invariant number density, \emph{viz.}
\begin{equation}
\label{eq:invariant number density}
    n_{\rm g,lc}(x) = i_j(\mathrm{d}V) = k\cdot j = n_g\,k \cdot u_g \;.
\end{equation}
Furthermore, by standard results from Riemannian geometry, if $\alpha$ is a $3$-form satisfying
\begin{equation}
  \mathrm{d}V \wedge \alpha = f\, \omega \;,
  \label{eq:alpha}
\end{equation}
then $\alpha = f \omtau$. Choosing $\alpha = \mathrm{d}U \wedge \mathrm{d}\zeta \wedge \mathrm{d}\bar{\zeta}$ yields $f \equiv \mathrm{i}$ and, therefore, the invariant volume form must be
\begin{equation}
\label{eq:invariant light cone volume form}
  \omtau = -\mathrm{i}~\mathrm{d}U \wedge \mathrm{d}\zeta \wedge \mathrm{d}\bar{\zeta} \;.
\end{equation}
This determines the ingredients of Eq.~\eqref{eq:galaxy number count} in terms of the null tetrad completely.\footnote{The simplest way to figure all these relations out is to consider a regular orthonormal tetrad $\theta^{\underline{a}}$ such that $\mathrm{d}U = \frac{1}{\sqrt{2}}(\theta^{\underline{0}}-\theta^{\underline{3}})$, $\mathrm{d}V = \frac{1}{\sqrt{2}}(\theta^{\underline{0}}+\theta^{\underline{3}})$ and $\mathrm{d}\zeta = \frac{1}{\sqrt{2}}(\theta^{\underline{1}}+\mathrm{i}\theta^{\underline{2}})$.} The $k,l$ system as well as the observer's past light cone is plotted in figure \ref{fig:cone}.

The null tetrad is thus far specified up to a fixed Lorentz transformation (which maps any null tetrad into another one). We have the freedom to choose the null vector field $k$, such that $k^\mu$ evaluated at the observer's location are the components of the photon wave vector, defined by the in-coming null geodesics from the galaxies in $\Sigtau$. This choice fixes 3 degrees of freedom. Clearly, $k$ is tangent to the past light cone $\Sigtau$ of the observer. Likewise, the null vector field $l$ is perpendicular to $\Sigtau$---as the latter is simply a level-curve $\{V = \textrm{const.}\}$ in $\mathbb{M}$---and, moreover, $l$ is normalised such that $k^\mu l_\mu = -1$.\footnote{In the literature (e.g. \cite{Penrose:1984uia,Stephanietal2003}), the definition is usually reversed, so that $k$ pertains to out-going null geodesics and $l$ to in-coming ones. We changed it here in order to keep $k^\mu$ as the four-velocity of the light ray from the source to the observer.} There are 2 degrees of freedom for the choice of $l$. We pick $l$ such that its spatial direction is opposite to that of $k$.
Finally, there is one more (irrelevant) degree of freedom in the choice of $m$, $\bar m$, for a total of 6 (corresponding to the 6 parameters of the Lorentz group). The null tetrad vectors and 1-forms are now fully specified.

In a local Lorentz frame $(t_L,r_L,\vartheta_L,\varphi_L)$ attached to the source position, these definitions amount to
\begin{equation}\label{eq: k and l in local frame}
\begin{aligned}
    k^a &= \frac{\beta}{\sqrt{2}}\big(1,-1,0,0\big) \\
    l^a &= \frac{1}{\beta\sqrt{2}}\big(1,1,0,0\big) \\
    m^a &= \frac{1}{r_L\sqrt{2}}\big(0,0,1,\frac{\mathrm{i}}{\sin\vartheta_L}\big) \;.
\end{aligned}
\end{equation}
The scalar $\beta$ corresponds to a boost in the plane $(k,l)$ that changes the frequency of the light beam. Using $\mathrm{d}U(l)=0$, $\mathrm{d}V(k)=0$ and the requirement that $\mathrm{d}U$ ($\mathrm{d}V$) be dual to $-l$ ($-k$) leads to
\begin{equation}\label{eq: U and V in local frame}
  \begin{aligned}
  \mathrm{d}U & = \frac{1}{\beta\sqrt{2}}\big(\mathrm{d}t_L - \mathrm{d}r_L\big) \\
  \mathrm{d}V & = \frac{\beta}{\sqrt{2}}\big(\mathrm{d}t_L + \mathrm{d}r_L\big) \\
  \mathrm{d}\zeta &= \frac{r_L}{\sqrt{2}}\big(\mathrm{d}\vartheta_L-\mathrm{i}\sin\vartheta_L\mathrm{d}\varphi_L\big)\;.
\end{aligned}
\end{equation}
Our choice of $\mathrm{d}U$ and $\mathrm{d}V$ ensures the conditions $\mathrm{d}U(k)= 1$ and $\mathrm{d}V(l)= 1$ or, equivalently, the null vectors $k$ and $l$ are normalised to $k\equiv \frac{\mathrm{d}}{\mathrm{d}U}$ and $l\equiv \frac{\mathrm{d}}{\mathrm{d}V}$ regardless of the value of $\beta$.

The family of integral curves---or congruence---of $k$ consists of the null geodesics $x^\mu(\lambda)$, where $\lambda$ is the affine parameter of the light rays. The condition $k^\mu \equiv \frac{\mathrm{d} x^\mu}{\mathrm{d}\lambda}$ implies $\mathrm{d}U\equiv \mathrm{d}\lambda$ and, therefore,
\begin{equation}
    -n_{\rm g,lc}(x) \mathrm{d}U = -n_g k_\mu u_g^\mu \mathrm{d}\lambda =: - n_g \mathrm{d}\ell\;.
\end{equation}
The minus sign ensures that $-n_{\rm g,lc}(x)\mathrm{d}U$ is positive ($k$ is incoming).
Let $\nvh$ be the sky direction as defined by the local observer.
In the local Lorentz frame where $k^\mu=E_s(1,-\nvh)$ ($E_s$ is the rest-frame energy of the emitted photon) and $u_g^\mu=\gamma_L(1,\vv_L)$ ($\vv_L$ is the source velocity in the local Lorentz frame), this becomes
\begin{equation}
\label{eq: proper length along light ray}
    -n_{\rm g,lc}(x)\mathrm{d}U = n_g E_s\gamma_L \big(1+\nvh\cdot\vv_L\big)\mathrm{d}\lambda
\end{equation}
Since $E_s\mathrm{d}\lambda = \mathrm{d}t_L = -\mathrm{d}r_L$ ($\mathrm{d}r_L<0$ for an incoming geodesic), we can also write
\begin{equation*}
    E_s\gamma_L\big(1+\nvh\cdot\vv_L\big)\mathrm{d}\lambda = -\gamma_L\big(\mathrm{d}r_L-\nvh\cdot\vv_L\mathrm{d}t_L\big)\;,
\end{equation*}
which shows that $\mathrm{d}\ell\equiv -k_\mu u_g^\mu\mathrm{d}\lambda>0$ is the proper length increment along the light beam measured in the galaxy rest frame. Therefore, this demonstrates that
\begin{equation}
    \label{eq:source physical volume form}
    \big(k\cdot u_g\big)\omega_\tau = \mathrm{i}\,\mathrm{d}\ell\wedge\mathrm{d}\zeta\wedge\mathrm{d}\bar\zeta
\end{equation}
is the physical volume form in the source rest frame.
Substituting this result into the definition \eqref{eq:galaxy number count} of $N(\Sigtau)$ yields
\begin{equation}
\label{eq:galaxy number counts forms final}
N(\Sigtau) = \int_{\Sigtau} \big(n_g \mathrm{d}\ell\big)\wedge\big(\mathrm{i}\,\mathrm{d}\zeta \wedge \mathrm{d}\bar{\zeta}\big)
\equiv \int_{\Sigtau} \big(n_g \mathrm{d}\ell\big)\wedge \sigma
\;.
\end{equation}

We have thus obtained a neat decomposition of the galaxy number count into a parallel component $n_g \mathrm{d}\ell$, and a transverse component with invariant area 2-form $\sigma$ that vanishes for vectors orthogonal to $m$ and $\bar m$. The latter defines a space-like hyper-plane (the polarisation plane) on which the separation vector $\xi^\mu=\bar\xi\, m^\mu+\xi\,\bar m^\mu$ between the fiducial and nearby geodesics is defined. For small $\xi$, the separation vector---or net deformation---can be obtained at any affine parameter value $\lambda$ by a linear transformation known as the Jacobi map \cite[e.g.,][]{Seitz:1994xf}. Alternatively, the deformation rate can be decomposed into optical scalars (expansion rate and shear) and evolved with the Sachs equations \cite[e.g.,][]{Sachs:1961zz}. Although we are chiefly interested in the weak regime of gravitational lensing, note that the existence of multiple images of a single source (i.e. caustics) implies that the Jacobi matrix is singular somewhere along the beam trajectory (in addition to the observer's position). As a result, the null vector $k$ would not be uniquely defined (since a single event on the past light cone will be connected to the observer with several distinct light rays). We will not consider such a situation here \cite[we refer the reader to the discussion in, e.g.,][]{Blandford:1986zz}, so that our basis $(k,l,m,\bar m)$ is always unambiguously defined.

One can repeat the procedure above and restrict $\Sigtau$ further to a given observed-redshift hyper-surface defined by the level curves of the observed redshift $\tilde z$. Denote this hyper-surface (\emph{qua} a sub-manifold of $\Sigtau$) by $\Sigtauz$.
Since, in practical applications, one is usually interested in redshift bins, it would be useful to find a function that, when integrated over $\mathrm{d}\tilde z$, gives $N$. Since proper length $\ell$ and observed redshift $\tilde z$ are in one-to-one correspondence along any given null geodesic, we immediately obtain
\begin{equation}\label{eqn: dN by dz 2.15}
  \frac{\mathrm{d} N}{\mathrm{d}\tilde z}(\Sigtauz) = \int_{\Sigtauz} n_g \bigg\lvert\frac{\mathrm{d}\ell}{\mathrm{d}\tilde z}\bigg\lvert \sigma\;.
\end{equation}
Finally, the data actually provides an estimate of the differential galaxy counts per unit redshift $\mathrm{d}\tilde z$ {\it and} observed unit area $\mathrm{d}\tilde \Omega =\mathrm{d}\!\cos\tilde\vartheta\wedge\mathrm{d}\tilde\varphi$ on the sky:
\begin{equation}
    \label{eq:dNdzdO}
    \frac{\mathrm{d}N}{\mathrm{d}\tilde{z}\mathrm{d}\tilde \Omega} = n_g \bigg\lvert\frac{\mathrm{d}\ell}{\mathrm{d}\tilde z}\bigg\lvert \det\!\mathcal D_o\;,
\end{equation}
where
\begin{equation}
\label{eq:detjacobimap}
\det\!\mathcal D_o \equiv \bigg\lvert\frac{\mathrm{i}\,\mathrm{d}\zeta\wedge\mathrm{d}\bar\zeta}{\mathrm{d}\!\cos\tilde\vartheta\wedge\mathrm{d}\tilde\varphi}\bigg\lvert
\end{equation}
is the determinant of the Jacobi map in the observer's frame.
This result corresponds to equation (4) of Ref.~\cite{ChallinorLewis2011}.
Eq.~\eqref{eq:detjacobimap} also makes clear that, since $\sigma$ is the proper beam area in the source rest frame and $\mathrm{d}\tilde\Omega$ is the observed area on the sky, $\det\mathcal{D}_o$ is nothing but the square of the angular-diameter distance to the source \cite[see the discussion in, e.g.,][]{Fleury:2015hgz}. Finally, our calculation implicitly assumes volume-limited surveys (of the past light cone $\Sigtau$). For flux-limited surveys, one should also include magnification bias \cite[e.g.,][]{gunn:1967,Moessner:1997qs,LoVerde:2007ke} through a flux limit in the definition of $N(\Sigtau)$.

\subsection{Expression in terms of inferred coordinates}
\label{sec:volume form in inferred coordinates}

We can now derive an expression for the observed galaxy count in terms of the observed redshift and sky position or, equivalently, as a function of the observer's inferred comoving coordinates $\tilde{x}^i=\chi(\tilde z)\nth^i$ on the past light cone $\Sigtau$, where $\chi(\tilde{z})$ is the inferred comoving distance to redshift $\tilde{z}$. These are the coordinates that a source, observed in direction $\ntvh$ with redshift $\tilde{z}$ would have had, had the Universe been an exact FLRW universe. The FLRW background assumed by the observer will generally not coincide with the true background, which can lead to distortions such as the Alcock-Paczynski effect \cite{Alcock:1979mp}. Here and henceforth, we will ignore this complication and assume that the background assumed by the observer coincides with the actual one.
In order to derive the effects of this (coordinate) projection on the observed galaxy over-density $\dgobs$, all we need to do is to express any of $\star j$, $\omega_\tau$ or $\sigma$ in terms of $\tilde{x}^i$.

To proceed, let $i:\Sigtau\to \mathbb{M}$ be the embedding map and $i^*$ the corresponding pull-back of exterior forms.
Starting from $\star j$ (equation \eqref{eq:galaxy number count Hodge}) written in some coordinate system $x^\mu$,
\begin{equation}
  \star j = \frac{1}{3!} \sqrt{|\det g|} \varepsilon_{\mu\nu\rho\lambda}\, j^\mu\, \mathrm{d}x^\nu\wedge\mathrm{d}x^\rho\wedge\mathrm{d}x^\lambda \;,
\end{equation}
the {\it restriction} of $\star j$ to the observer's past light cone, parameterised with the coordinates $\tilde{x}^i$, precisely is the pull-back of $\star j$ to $\Sigtau$. Explicitly,
\begin{equation}\label{eq: star j resticted}
\begin{aligned}
    i^*\big(\star j\big) &= \frac{1}{3!} \sqrt{|\det g|} \varepsilon_{\mu\nu\rho\lambda}\, j^\mu(\tvx)\, i^*\!\big(\mathrm{d}x^\nu\big)\wedge i^*\!\big(\mathrm{d}x^\rho\big)\wedge i^*\!\big(\mathrm{d}x^\lambda\big) \\
    &=\sqrt{|\det g|} n_g(\tvx) \varepsilon_{\mu\nu\rho\lambda}\, u_g^\mu(\tvx)\,
    \frac{\partial x^\nu}{\partial \tilde{x}^1}\frac{\partial x^\rho}{\partial \tilde{x}^2}\frac{\partial x^\lambda}{\partial \tilde{x}^3}\mathrm{d}\tilde x^1\wedge\mathrm{d}\tilde x^2\wedge\mathrm{d}\tilde x^3
\end{aligned}
\end{equation}
where $\tilde x^i$ are the local coordinates on $\Sigtau$ and the notation $n_g(\tvx)$ designates $n_g\big(x^\mu(\tilde x^i)\big)$. Here, we have used
the relation
\begin{equation}
\begin{aligned}
        i^*\!\big(\mathrm{d}x^\nu\big)\wedge i^*\!\big(\mathrm{d}x^\rho\big)\wedge i^*\!\big(\mathrm{d}x^\lambda\big) &= \frac{\partial x^\nu}{\partial \tilde{x}^i}\frac{\partial x^\rho}{\partial \tilde{x}^j}\frac{\partial x^\lambda}{\partial \tilde{x}^k}\mathrm{d}\tilde x^i\wedge\mathrm{d}\tilde x^j\wedge\mathrm{d}\tilde x^k \\
        &= 3! \frac{\partial x^\nu}{\partial \tilde{x}^1}\frac{\partial x^\rho}{\partial \tilde{x}^2}\frac{\partial x^\lambda}{\partial \tilde{x}^3}\mathrm{d}\tilde x^1\wedge\mathrm{d}\tilde x^2\wedge\mathrm{d}\tilde x^3 \;.
\end{aligned}
 \end{equation}
Since $\star j = n_g \big(k\cdot u_g\big)\omtau$, we can immediately read off that
\begin{equation}
\label{eq:k.u omega tau}
  i^*\big[\big(k\cdot u_g\big)\omtau\big] = \sqrt{|\det g|}\, \varepsilon_{\mu\nu\rho\lambda}\, u_g^\mu(\tvx)\,
  \frac{\partial x^\nu}{\partial \tilde{x}^1}\frac{\partial x^\rho}{\partial \tilde{x}^2}\frac{\partial x^\lambda}{\partial \tilde{x}^3}
  \,\mathrm{d}\tilde x^1\wedge\mathrm{d}\tilde x^2\wedge\mathrm{d}\tilde x^3
  \;,
\end{equation}
and, substituting into \eqref{eq:galaxy number count Hodge}, we obtain
\begin{equation}\label{eq:galaxy number counts coordinates}
  N(\Sigtau) = \int_{\Sigtau} \mathrm{d}\tilde x_1\mathrm{d}\tilde x_2\mathrm{d}\tilde x_3\, \sqrt{\abs{\det{g}}}n_g(\tvx)\eps_{\mu\nu\rho\lambda}u_g^\mu(\tvx)\frac{\partial x^\nu}{\partial \tilde{x}^1}\frac{\partial x^\rho}{\partial \tilde{x}^2}\frac{\partial x^\lambda}{\partial \tilde{x}^3},
\end{equation}
where the integral now runs over $\Sigtau$ as parameterised by $\tilde x^i$. This matches the expression given in Ref.~\cite{Choquet-Bruhat2009}. An alternative derivation that explicitly relies on the null tetrad, that is, the light-cone structure of the volume form, can be found in Appendix \ref{sec:omega_tau in terms of inferred coordinates}.

In order to define the observed galaxy number density $\tilde{n}_g$ on the past light cone $\Sigtau$ in terms of the inferred comoving coordinates $\tilde x^i$, we express $N(\Sigtau)$ as
\begin{equation}\label{eq:number counts in inferred coordinates}
    N(\Sigtau) \equiv \int_{\Sigtau}\tilde{n}_g(\tvx)\,\tomtau\,.
\end{equation}
Here, the basic 3-form of the local coordinates on $\Sigtau$ is
\begin{equation}\label{eqn: inferred 3-volume form}
\tomtau\equiv\sqrt{|\det\overline{g}(\tilde{a})|}\,\mathrm{d}\tilde x^1\wedge\mathrm{d}\tilde x^2\wedge\mathrm{d}\tilde x^3\;,
\end{equation}
and is defined with the scale factor $\tilde{a}\equiv 1/(1+\tilde{z})$, with $\overline{g}(\tilde{a})$ being the FLRW background metric evaluated at the inferred scale factor.
The inferred coordinates $\tilde x^i$ may, for example, be decomposed into the spherical coordinates $\tilde{x}^i=(\tilde\chi,\tilde\vartheta,\tilde\varphi)$ (so that $\tomtau=\tilde{a}^3\tilde\chi^2\,\mathrm{d}\tilde{\chi} \wedge\mathrm{d}\tilde\Omega$ for a spatially flat Universe\footnote{The inferred angles $(\tilde\vartheta,\tilde\varphi)$ can be used to define the 1-forms $\mathrm{d}\tilde\zeta = \frac{\tilde\chi}{\sqrt{2}}(\mathrm{d}\tilde\vartheta+\mathrm{i}\sin\tilde\vartheta\mathrm{d}\tilde\varphi)$,
$\mathrm{d}\bar{\tilde\zeta} = \frac{\tilde\chi}{\sqrt{2}}(\mathrm{d}\tilde\vartheta-\mathrm{i}\sin\tilde\vartheta\mathrm{d}\tilde\varphi)$ (transverse to the observed sky direction $\ntvh$) and, thereby, express $\mathrm{d}\tilde\Omega$ as $\mathrm{d}\tilde\Omega=\frac{\mathrm{i}}{\tilde\chi^2}\mathrm{d}\tilde\zeta\wedge\mathrm{d}\bar{\tilde\zeta}$.}) or into $\tilde x^i=(\tilde z,\ntvh)$ (so that $\tomtau\propto \mathrm{d}\tilde z\wedge\mathrm{d}\tilde\Omega$), \emph{etc.}
Both $\tomtau$ and $i^*(\omtau)$ are top-forms (i.e. 3-forms) on $\Sigtau$ and, as such, are related by a multiplicative function which is nothing but a Jacobian (see equation \eqref{eq:linkingomtautotomtau}).

\section{non-linear, gauge-invariant observed galaxy over-density}
\label{sec:non-linear delta g}

We defined the galaxy number count in terms of the comoving coordinates $\tilde x^i$ inferred by the observer above. Note that we have neither made any assumption about the coordinates chosen to describe $\mathbb{M}$, nor about the smallness of the perturbations around the FLRW background. In this section, we apply our results to derive an expression for the  non-linear galaxy over-density $\dgobs$, inferred by $\mathcal{O}$ in terms of gauge-invariant, non-linearly defined observable quantities---a task that can be implemented with the ``cosmic ruler'' observables introduced in Refs.~\cite{SchmidtJeong2012,JeongSchmidt2014}. We start from Eq.~\eqref{eq:source physical volume form}, and carry out the calculation in as general a manner as possible.
In Appendix \ref{sec:alternative}, we present a more standard derivation at linear order in perturbations starting from Eq.~\eqref{eq:galaxy number counts coordinates}, similar to that of Refs. \cite{Jeong:2011as,SchmidtJeong2012}.
The final result agrees with that of this section in the appropriate limit.

\subsection{Gauge-invariant expression}
\label{sec:gauge invariant expression}

We begin with the $3$-form $(k\cdot u_g)\omtau = \mathrm{i}\,\mathrm{d}\ell \wedge \mathrm{d}\zeta \wedge \mathrm{d}\bar{\zeta}$ written in terms of the inferred coordinates $\tilde x^i$, that is, with its pull-back to $\Sigtau$.
Rather than attempting a brute force evaluation of Eq.~\eqref{eq:k.u omega tau}, we simplify the analysis by considering a standard ruler with a physical size $r_0(\tau)$ as measured in its rest frame.

Consider a given observed galaxy, and a set of three orthogonal equal-length imaginary rods (rulers) $\{\bm{r}_a,\bm{r}_b,\bm{r}_c\}$.
Specifically, the end points of each ruler correspond to two observers comoving with the galaxies, and which at an arbitrary fixed proper time $\tau_s$ have a proper physical separation of $r_0$. In the case at hand, this time will be the proper time at which the observed galaxy's light was emitted.
We assume for now that we have a way of determining the emission time and hence the physical scale $r_0$.

\begin{figure}[t]
    \centering
    \includegraphics[width=0.8\textwidth,trim=0cm 13cm 0cm 0cm,clip=true]{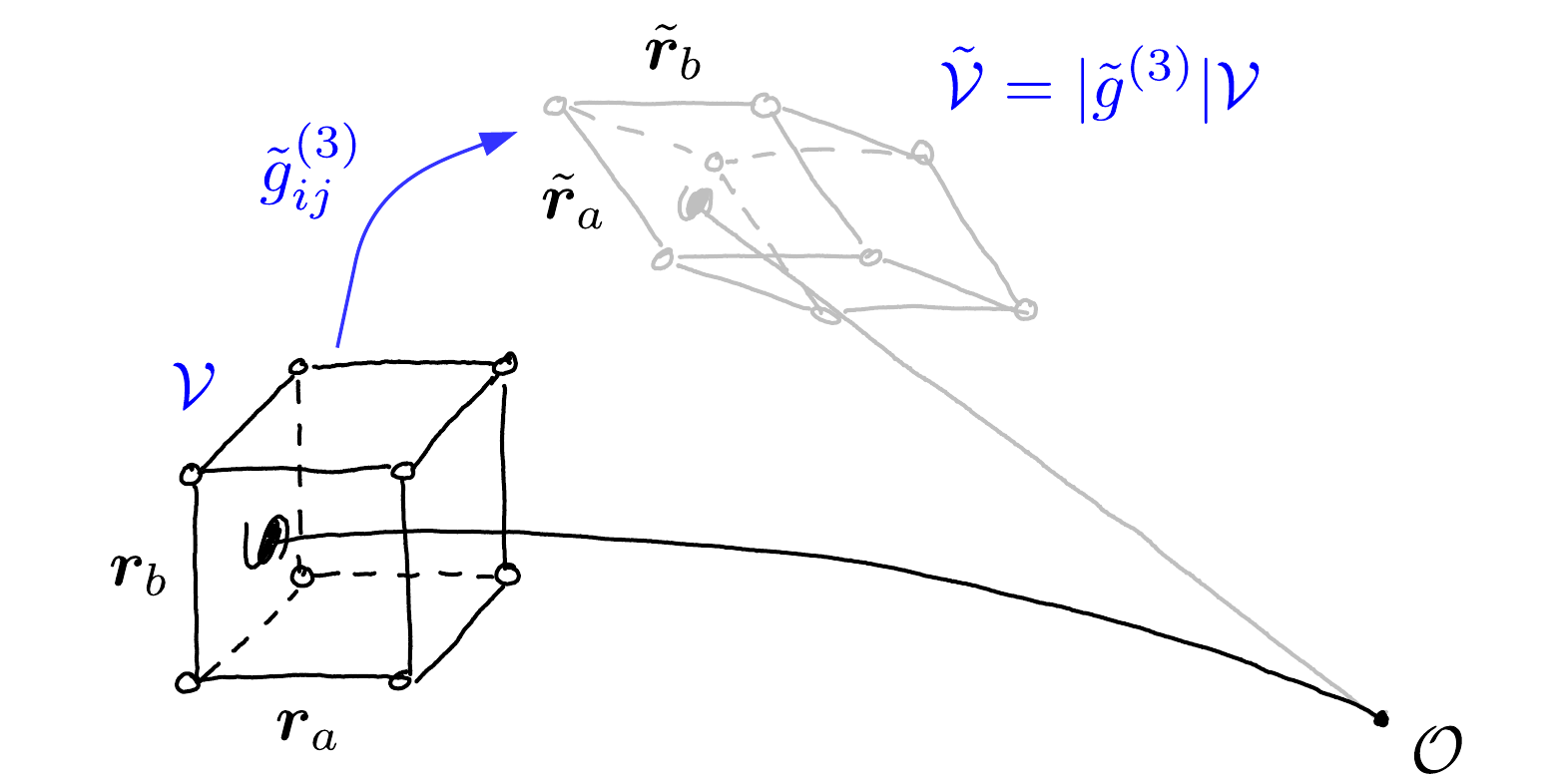}
    \caption{Sketch of how ruler perturbations describe the volume distortion by projection effects. A cube of infinitesimal size centred on an observed galaxy at position $\bm{x}$ is distorted (sheared and stretched) into a more general parallelepiped around the  observationally inferred position $\tilde{\bm{x}}$. This distortion is described by the metric \refeq{tildeg}. The apparent volume is correspondingly distorted by a factor $\det\tilde g^{(3)}$.}
    \label{fig:rulers}
\end{figure}

Now imagine light bulbs attached to the ends of these rods, which form an imaginary cube (see figure \ref{fig:rulers}). By following the photon geodesics from the end-points of the rulers to the distant observer, we obtain the apparent size and orientations of each ruler as determined by the observer. In the limit of infinitesimal rulers, i.e. of rulers spanning small angles and redshift intervals as seen by the distant observer, the rest-frame cube is distorted into a parallelepiped surrounding the observed position of the galaxy. For the purpose of computing the observed galaxy density, one can always choose the physical size of the imaginary rulers small enough to satisfy this criterion.

We can now determine the true and apparent physical volume of the parallelepipeds spanned by these rods. The physical rest-frame density of the galaxies is then given by the number of galaxies within the actual parallelepiped (cube) divided by its volume, while the observed number density is given by the same number of galaxies divided by the volume of the \emph{apparent} parallelepiped spanned by the rulers as seen by the observer, $\tilde{\bm{r}}_a,\tilde{\bm{r}}_b,\tilde{\bm{r}}_c$.
More precisely, the physical volume spanned by the rulers is given by
\begin{equation}\label{eqn: equation 3.1}
\big(k\cdot u_g\big)\omtau(\bm{r}_{0,a},\bm{r}_{0,b},\bm{r}_{0,c}) \propto \ell \; \sigma \propto r_0^3(\tau_s).
\end{equation}
On the other hand, the apparent volume spanned by the rulers as seen by the distant observer is given by the apparent volume form $\tomtau$. The two are related by the distortion induced in the apparent size of the rulers by the light propagation. This can be described by the Jacobian of an effective 3D metric parameterised by the ruler perturbations, as we will describe below. Thus,
\begin{equation}
\label{eq:linkingomtautotomtau1}
    i^*\big[\big(k\cdot u_g\big)\omtau\big] \stackrel{\mbox{fixed $\tau_s$}}{=} \mathcal  \det \big(\tilde g^{(3)}\big) \;\tomtau \;,
\end{equation}
where we have noted that this relation holds if $\tomtau$ is evaluated at the
source's proper time of emission. In practice, the distant observer evaluates $\tomtau$ at the observed redshift, leading to an additional factor we will discuss below.

In the limit of infinitesimal rulers, which we can assume given the reasoning above, the mapping from proper physical length $r_0$ to observationally inferred length $\tilde r$ can be decomposed as
\begin{equation}\label{eq:ruler definitions}
  \left(\begin{array}{c}
          r_{0,\parallel} \\[2pt]
          r_{0,\perp}^1 \\[2pt]
          r_{0,\perp}^2
        \end{array}\right) = \left(\begin{array}{ccc}
                                     1-\mathcal{C} & -\mathcal{B}_1 & -\mathcal{B}_2 \\[2pt]
                                     -\mathcal{B}_1 & 1-\mathcal{A}_{11} & -\mathcal{A}_{12} \\[2pt]
                                     -\mathcal{B}_2 & -\mathcal{A}_{21} & 1-\mathcal{A}_{22}
                                   \end{array}\right)\left(\begin{array}{c}
                                                             \tilde{r}_{\parallel} \\[2pt]
                                                             \tilde{r}_{\perp}^1 \\[2pt]
                                                             \tilde{r}_{\perp}^2
                                                           \end{array}\right).
\end{equation}
Here, $\tilde r_\parallel \equiv \tilde r^i \hat{n}_i$ and $\tilde{r}_\perp^i = (\delta^{ij} - \hat n^i \hat n^j) \tilde r_j$ are the line-of-sight and transverse components of the observed ruler, and likewise for $r_{0,\parallel}$, $r_{0,\perp}^i$. The coefficients $\mathcal C$, $\mathcal B_i$ and $\mathcal A_{ij}$ define the distortion on, respectively, the radial-radial, the radial-tangential, and the tangential-tangential planes \citep[see][for details]{SchmidtJeong2012,JeongSchmidt2014}. Namely, the mapping Eq. \eqref{eq:ruler definitions}
is a generalisation to finite perturbations of the cosmic rulers introduced in \cite{SchmidtJeong2012}, who worked at linear order in perturbations around FLRW. To see this, consider the perturbation to the length of the ruler, $\tilde r = ( \tilde{\bm{r}}^2 )^{1/2}$, in the limit of small perturbations $\mathcal C$, $\mathcal B_i$ and $\mathcal A_{ij}$. Using \refeq{ruler definitions} yields
\begin{equation}
  \frac{\tilde{r} - r_0}{\tilde r} = \mathcal C \frac{\tilde{r}_\parallel^2}{\tilde r^2} + 2\mathcal B_i \frac{\tilde{r}_\parallel\tilde{r}_\perp^i}{\tilde r^2} + \mathcal A_{ij} \frac{\tilde{r}_\perp^i\tilde{r}_\perp^j}{\tilde r^2}\;,
    \label{eq:localmetric}
\end{equation}
which is Eq.~(31) of \cite{SchmidtJeong2012}, except for a factor of 2 in the definition of $\mathcal{B}_i$, which we adopt here for consistency with the remaining metric components.

We have written Eq. \eqref{eq:ruler definitions} as a mapping that yields the proper physical length of an observed ruler $\tilde{\bm{r}}$. This immediately leads to the natural definition of the local metric $\tilde{g}^{(3)}_{ij}$ relating the observer's inferred coordinates to proper physical distances in the source rest frame. Expressing $\tilde{\bm{r}}_\perp$ in terms of the circular polarisations $m$ and $\bar m$ (\refeq{nulltetrad}) and being attentive to the sign,
we have
\begin{equation}
  \tilde g^{(3)} = I_{3\times3} - \left(\begin{array}{ccc}
          \mathcal{C} & \mathcal{B}_m & \mathcal{B}_{\bar{m}} \\
          \mathcal{B}_m & \mathcal{A}_{mm} & \mathcal{A}_{m\bar{m}} \\
          \mathcal{B}_{\bar{m}} & \mathcal{A}_{\bar{m}m} & \mathcal{A}_{\bar{m}\bar{m}}
  \end{array}\right)\;.
  \label{eq:tildeg}
\end{equation}
Here, $I_{3\times3}$ is the $3\times3$ identity matrix.
While Ref.~\cite{SchmidtJeong2012} derived linear-order expressions for $\mathcal{C,B,A}$, their definition is not restricted to small perturbations around an FLRW spacetime; all we have assumed here are infinitesimal rulers. Thus, the metric \refeq{tildeg} provides an exact, non-perturbative relation between the intrinsic volume in the rest frame of the galaxy and the volume derived from the inferred coordinate of the observer, both at fixed proper time of emission.
Using the components in \refeq{tildeg}, the determinant in \refeq{linkingomtautotomtau1} becomes
\begin{equation}
\label{eq:determinant explicitly}
\begin{aligned}
    \det\big(\tilde g^{(3)}\big) & = (1-\mathcal{C})(1-\mathcal{A}_{mm})(1-\mathcal{A}_{\bar{m}\bar{m}}) - \mathcal{B}_m\mathcal{B}_{\bar{m}}(\mathcal{A}_{m\bar{m}} + \mathcal{A}_{\bar{m}m}) \\ & \quad -
    (\mathcal{B}_{\bar{m}}^2 + \mathcal{B}_m^2)
    + \mathcal{B}_{\bar{m}}^2\mathcal{A}_{mm}
    + \mathcal{B}_m^2\mathcal{A}_{\bar{m}\bar{m}}
    -\mathcal{A}_{m\bar{m}}\mathcal{A}_{\bar{m}m}(1-\mathcal{C}) \;.
\end{aligned}
\end{equation}

Next, we need to take into account that the distant observer does not have access to the proper time of emission $\tau_s$ directly, but rather the observed redshift of the source.
This leads to an additional factor
\begin{equation}
  \mathcal{J}_{\mathcal{T}} = \left(\frac{r_0(\tau_s)}{r_0(\overline{\tau}(\tilde a))}\right)^3 \;,
  \label{eq:JT}
\end{equation}
where $\overline{\tau}(\tilde a)$ is computed from the background proper time-redshift relation.
In other words, $r_0(\overline{\tau}(\tilde a))$ is the expected infinitesimal ruler size given the observed redshift, $\tilde a = 1/(1+\tilde z)$, which arises because the distant observer computes volumes based on the fiducial background relation, while
$r_0(\tau_s)$ is the actual size of the ruler at the proper time of emission.
In order to evaluate \refeq{JT}, we need to know how the ruler size evolves.\footnote{Strictly speaking, we should write \refeq{JT} in terms of the product $r_{0,a}r_{0,b}r_{0,c}$, and allow for a different time dependence of each ruler component. The end result, \refeq{r0oftau}, is the same however.}
The statement that the end points of the ruler are comoving with the galaxies implies that the time evolution of the volume spanned by the rulers is given by the local velocity divergence,
\begin{equation}
  \frac{{\rm d} \ln r_0^3(\tau_s)}{{\rm d}\tau_s} = \nabla_\mu u_g^\mu(\tau_s) \equiv \veldiv(\tau_s)
  \;,
  \label{eq:r0oftau}
\end{equation}
where we have introduced the four-velocity divergence as $\veldiv$.
At zeroth order in perturbations around an FLRW background, $r_0(\tau_s) \propto a(\tau_s)$; that is, in an unperturbed background the physical ruler size simply follows the background expansion. By integrating \refeq{r0oftau} along the galaxy geodesics (the integral curves of $u_g$), we can then evaluate \refeq{JT} fully non-linearly. Notice that the absolute size of the ruler is an integration constant in \refeq{r0oftau} but drops out of \refeq{JT}.
We turn to the derivation of $\mathcal J_{\mathcal{T}}$ in the following subsection (Sec.~\ref{sec:non-linearT}).
Finally, multiplying this factor to \refeq{linkingomtautotomtau1}, we arrive at
\begin{equation}
  \label{eq:linkingomtautotomtau}
  i^*\big[\big(k\cdot u_g\big)\omtau\big] = \mathcal J_{\mathcal T} \; \det\!\big(\tilde g^{(3)}\big) \;\tomtau \;.
\end{equation}
To summarise, the right-hand side here contains a product of three terms, capturing $(1)$ the fact that the proper time and, hence, the scale factor $a(\tau_s)$ differs from the observationally inferred one $\tilde a = 1/(1+\tilde z)$, leading to a factor $\mathcal J_{\mathcal{T}}$; $(2)$ the determinant $\det\!\big(\tilde g^{(3)}\big)$ of the spatial metric relating volumes on the actual past light cone of the observer to the apparent one via ruler distortions; $(3)$ the volume form $\tomtau$ on the \emph{apparent} past unperturbed light cone. The volume form on the left-hand side depends on the galaxy four-velocity $u_g$ (which is equal to the matter four-velocity on large scales). This dependence is encoded on the right-hand side both in the factor $\mathcal J_{\mathcal T}$, since the time
evolution of the rulers is given by the four-velocity divergence, and in the Jacobian $\det(\tilde g^{(3)})$, since the metric $\tilde g^{(3)}$ is defined on slices orthogonal to $u_g$.

We note that our reasoning remains valid even in the strong-lensing regime, where a single source is mapped to multiple images \cite{weinberg:1976,bartelmann/schneider:2001}. In that case, however, one needs to sum \refeq{linkingomtautotomtau} over the individual images. Further, in the application to gravitational lensing, finite-ruler or equivalently finite-source-size effects in general need to be included \cite{2006MNRAS.365.1243D,PhysRevLett.119.191101}. In the remainder of the paper, we disregard the strong-lensing regime.

\subsection{non-linear definition of $\mathcal T$}\label{sec:non-linearT}

Ref.~\cite{JeongSchmidt2014} has introduced the {\it cosmic clock} observable $\mathcal{T}_a$ in terms of the logarithmic difference between the scale factor $\at(\tau_s)$ evaluated at a constant source proper time, and the scale factor $\tilde a = 1/(1+\tilde z)$ inferred from the observed redshift,
\begin{equation}
\label{eq:defTa}
  \mathcal{T}_a \equiv \ln \left(\frac{\at(\tau_s)}{\tilde a}\right)\;.
\end{equation}
We use the symbol $\at(\tau)\equiv a\big(\bar\eta(\tau)\big)$ to emphasise that $\at$ is parameterised in terms of $t$, as opposed to $a(\eta)$.
Note that $\mathcal{T}_a$ vanishes in the limit where the source overlaps with the observer (i.e. $\tilde z\to 0$) since $\at(\tau_o)\equiv 1$.
We also emphasise that $\mathcal{T}_a$ is a genuine observable since both $\tau_s$ and $\tilde{z}$ are observable quantities, and the function $\at(\tau_s)$ is fully defined in terms of a solution to the Friedmann equations (up to an irrelevant global re-scaling). Thus defined, $\mathcal{T}_a$ is an observable and, therefore, fully gauge-invariant.

Ref.~\cite{JeongSchmidt2014} denoted Eq.~\eqref{eq:defTa} as $\mathcal{T}$, working at linear order in perturbations. Here, we extend this relation to find a fully non-linear definition $\mathcal{T}_a$.
In this section, we consider a general time-evolving ruler scale $r_0(\tau)$. For the purposes of deriving the physical galaxy number density as explained above, the sensible choice corresponds to a local ruler which evolves according to equation \eqref{eq:r0oftau}. We will thus specialise to this case in the remainder of the paper.

Consider now an infinitesimal physical volume $V_0\equiv r_0^3(\tau_s)$ in the source rest frame. For convenience, we now parameterise the time dependence using the scale factor, via the scale-factor-proper-time relation in the background.
Correspondingly, we denote $V_0(\tilde a)\equiv V_0(\overline\tau(\tilde a))$ in the following. We can generally write
\begin{equation}
    \begin{aligned}
    V_0(\tau_s) &= V_0(\tilde a) + \mathcal{T}_a\frac{\mathrm{d}V_0}{\mathrm{d}\ln \tilde{a}}(\tilde a) + \frac{\mathcal{T}_a^2}{2}\frac{\mathrm{d}^2V_0}{\mathrm{d}(\ln \tilde{a})^2}(\tilde a) + \ldots \\
    &= \left[\exp\!\left(s\frac{\mathrm{d}}{\mathrm{d}\ln \tilde{a}}\right)V_0(\tilde a)\right]_{s = \mathcal{T}_a}\;.
    \label{eq:V0_Ta}
  \end{aligned}
\end{equation}
The Jacobian $\mathcal J_{\mathcal{T}}$ is then
\begin{equation}
    \label{eq:dV0 by dlna}
  \mathcal{J}_{\mathcal{T}} = V_0(\tilde a)^{-1} \left[\exp\!\left(s\frac{\mathrm{d}}{\mathrm{d}\ln \tilde{a}}\right)V_0(\tilde a)\right]_{s = \mathcal{T}_a}  \;,
\end{equation}
i.e. the exponentiation of an infinitesimal translation, acting on $V_0$.\footnote{One can also view the operator $J \equiv \exp\left(\mathcal{T}_a\frac{\mathrm{d}}{\mathrm{d\ln \tilde{a}}}\right)$ as a 1-dimensional operator acting on the 1-dimensional function-space spanned by the function $V_0(\tilde{a})$, in which case the Jacobian is its determinant, $\exp \mathrm{tr} \ln J$, where the eigenvalue of $J$ is simply, by definition, $V_0(\at(\tau_s))/V_0(\tilde{a})$.}

A second option for defining the cosmic clock observable non-linearly, which we denote by $\mathcal{T}_r$, consists of defining $\mathcal{T}$ through the ratio of ruler sizes measured on a constant source proper time hyper-surface, and on a constant observed redshift hyper-surface. Namely,
\begin{equation}
    \frac{r_0\big(\at(\tau_s)\big)}{r_0(\tilde a)} \equiv 1 + \mathcal{T}_r\,\frac{\mathrm{d}\ln r_0}{\mathrm{d}\ln \tilde{a}}\!(\tilde a) \;,
\end{equation}
whence the Jacobian is simply
\begin{equation}
  \mathcal{J}_{\mathcal{T}} = \left(1+ \frac{1}{3}\frac{\mathrm{d}\ln V_0}{\mathrm{d}\ln \tilde{a}}\!(\tilde a)\,\mathcal{T}_r\right)^3 \;.
\end{equation}
While the latter Jacobian appears simpler than the former,
its definition relies on a particular evolution law of the ruler $r_0(\tau)$.
Hence, we will continue to work with the former non-linear definition, $\mathcal{T}_a$. Note again however that $\mathcal{T}_a$ and $\mathcal{T}_r$ are equal at linear order.

$\mathcal{T}_a$ has another nice interpretation: suppose one wishes to compute the perturbation $\delta z$, to the observed redshift \cite{Fanizzaetal2018}, defined by
\begin{equation}
    1+\tilde{z} \equiv \frac{\at(\tau_o)}{\at(\tau_s)}(1+\delta z) = \frac{1}{\at(\tau_s)}(1+\delta z).
\end{equation}
Then, by definition of $\tilde{a} = (1+\tilde{z})^{-1}$, we find
\begin{equation}
    \delta z = e^{\mathcal{T}_a} - 1,
\end{equation}
at all orders in perturbation theory, which implies that the redshift perturbation is a function of $\mathcal{T}$ only, and automatically gauge-invariant. To compute it for a given $\tilde{z}$ to a given order, one needs to know the perturbation to the proper time at the appropriate order, and insert it into $\at(1+\tilde{z})$.

\subsection{Non-linear and linear galaxy over-density}

Having defined the cosmic clock (or time ``ruler'') ${\cal T}$
non-linearly, a comparison between the pull-back $i^*(\star j)$,
\begin{equation}
    i^*\big[n_g(x)\big(k\cdot u_g\big)\omtau\big] =
    n_g(\tilde x) \mathcal J_{\mathcal T} \det\!\big(\tilde g^{(3)}\big) \,\tomtau\;,
\end{equation}
and the inferred galaxy counts $\tilde n_g(\tilde x)\,\tomtau$ yields a \emph{fully non-linear} expression for the observed galaxy over-density $\dgobs$ in terms of gauge-invariant ruler quantities $\mathcal{C}$, $\mathcal{B}_i$, $\mathcal{A}_{ij}$ and $\mathcal{T}_a$.

To see this, we write the inferred galaxy density $\tilde n_g(\tvx)$ as
\begin{equation}
    \label{eq:def of observed delta_g}
    \tilde n_g(\tvx) = \bar n_g(\tilde z)\Big[1+\dgobs(\tvx)\Big]\;,
\end{equation}
where $\bar n_g(\tilde z)$ is the mean physical galaxy number density at observed redshift $\tilde z$. On defining perturbations to the pull back of the galaxy rest frame density relative to the same constant observed redshift hyper-surface,
\begin{equation}
    n_g(\tvx) = \bar n_g(\tilde z) \Big[1+\delta_g^\textrm{or}(\tvx)\Big]\;,
\end{equation}
and on equating the integrand of Eq.~\eqref{eq:galaxy number counts forms final} with Eq.~\eqref{eq:number counts in inferred coordinates}, we straightforwardly obtain
\begin{equation}
    \label{eq:over density non linear}
  1 + \dgobs(\tvx) = \Big[1+ \delta_g^\textrm{or}(\tvx)\Big]\, \det\!\big(\tilde g^{(3)}\big)\, V_0(\tilde a)^{-1} \left[\exp\!\left(s\frac{\mathrm{d}}{\mathrm{d}\ln \tilde{a}}\right)V_0(\tilde a)\right]_{s = \mathcal{T}_a}\;.
\end{equation}
We can transform from \emph{constant observed redshift} (superscript or) slices to constant \emph{source proper time} (superscript pt) slices by making use of the definition of $\mathcal{T}_a$, specifically \refeq{V0_Ta}:
\begin{equation}
n_g(\tvx) = \bar n_g(\tilde z) \Big[1+\delta_g^{\rm or}(\tvx)\Big]
= \left[\exp\!\left(s \frac{\partial}{\partial\ln a}\right)\bigg(
\bar n_g(\tau(\tilde z)) \Big[1+\delta_g^{\rm pt}(\tvx)\Big]\bigg)\right]_{s=\mathcal{T}_a}\;.
\label{eq:or-to-pt}
\end{equation}
Theoretical predictions for the rest-frame galaxy density, for example using a bias expansion, are naturally given on constant-proper-time slices.
Likewise, expanding out the various Jacobians in Eq.~\eqref{eq:over density non linear} is carried out by using the fact that the rest-frame infinitesimal ruler's volume changes according to
\begin{equation}
    \frac{1}{V_0(\tau_s)}\frac{\mathrm{d}V_0}{\mathrm{d}\tau_s} = \veldiv(\tau_s)\;.
\end{equation}
Consequently
\begin{align}
    \frac{\mathrm{d}V_0}{\mathrm{d}\ln \tilde{a}} & = \frac{\mathrm{d}V_0}{\mathrm{d}\tau}\left(\frac{\mathrm{d}\overline\tau}{\mathrm{d} \ln a}\right)_{\tilde{a}} \\ &
    = V_0(\tilde a) \veldiv(\tilde a) \tilde{H}^{-1} \;,
\end{align}
where $\tilde{H}$ is the background expansion rate evaluated at $\tilde a$.
As a result, Eq.~\eqref{eq:dV0 by dlna} yields
\begin{equation}\label{eqn: equation 3.25}
  \mathcal{J}_\mathcal{T} = 1 + \mathcal{T}_a \frac{\theta(\tilde a)}{\tilde{H}} + \frac{\mathcal{T}_a^2}{2}\left(
  \left[\frac{\theta(\tilde{a})}{\tilde{H}}\right]^2
  +  \frac{\mathrm{d}}{\mathrm{d}\ln \tilde{a}}\frac{\theta(\tilde{a})}{\tilde{H}}\right) + \ldots \;.
\end{equation}
At linear order in the perturbations, $\mathcal{J}_\mathcal{T} \approx 1+3\mathcal{T}_a$, since $\mathcal{T}_a$ is linear order, and $\veldiv = 3 \tilde H$ at zeroth order. By equation \eqref{eq:r0oftau}, $V_0(\tilde{a})$ does not appear in the fully expanded expression for $\mathcal{J}_{\mathcal{T}}$. In conjunction with Eq.~\eqref{eq:determinant explicitly}, this implies that the observed, non-linear galaxy over-density can be expressed as
\begin{align}\label{eq:delta non-linear rulers}
    \frac{1 + \dgobs(\tvx)}{1+ \delta_g^\textrm{or}(\tvx)} & = \mathcal{J}_{\mathcal{T}}
    (1-\mathcal{C})(1-\mathcal{A}_{mm})(1-\mathcal{A}_{\bar{m}\bar{m}}) - \mathcal{J}_{\mathcal{T}}\Big[ \mathcal{B}_m\mathcal{B}_{\bar{m}}(\mathcal{A}_{m\bar{m}} + \mathcal{A}_{\bar{m}m}) \\ &\qquad +
    (\mathcal{B}_{\bar{m}}^2 + \mathcal{B}_m^2)
    - \mathcal{B}_{\bar{m}}^2\mathcal{A}_{mm}
    - \mathcal{B}_m^2\mathcal{A}_{\bar{m}\bar{m}}
    +\mathcal{A}_{\bar{m}m}\mathcal{A}_{m\bar{m}}(1-\mathcal{C})
    \Big] \nonumber \;.
\end{align}
Equations \eqref{eq:over density non linear} and \eqref{eq:delta non-linear rulers} are the main results of this section.

Besides the galaxy density contrast $\delta_g^{\rm or}$ on the constant-observed-redshift hyper-surface, which is by itself an observable, the projection effects in equation \eqref{eq:delta non-linear rulers} consist only of observables such as the line-of-sight and transverse cosmic rulers (${\cal C}$, ${\cal B}_i$, ${\cal A}_{ij}$) and the cosmic clock (${\cal T}$).
It is a much simpler task to evaluate $\mathcal{T}, \mathcal{C}$, $\mathcal{B}_i$, $\mathcal{A}_{ij}$ to the required order in perturbation theory than to do so directly for the entire $\dgobs$. This decomposition into simpler terms opens up a new avenue of handling the observed galaxy density contrast at non-linear order, which is essential for modelling, for example, the observed galaxy bispectrum on large scales.

Finally, we can specialise to linear order in perturbations. The expansion $\det (I+A) \approx 1 + \tr A$ gives
\begin{equation}
    \label{eq:over density linear}
    \dgobs(\tvx) = -\mathcal{C}-\mathcal{M} + 3
    \mathcal{T} + \delta^\textrm{or}_g(\tvx) \;,
\end{equation}
where $\mathcal{M}\equiv \mathcal{A}_{mm}+\mathcal{A}_{\bar m\bar m}$ is the magnification at linear order. Moreover, the transformation to constant source-proper-time slices \refeq{or-to-pt} becomes
\begin{equation}\label{eq:delta or}
  \delta_g^\textrm{or} = \delta_g^\textrm{pt} + \frac{\mathrm{d}\ln \bar{n}_g}{\mathrm{d}\ln \tilde{a}}\!(\tilde a)\, \mathcal{T}
\end{equation}
in linear theory, for which the notation $\mathcal{T}\equiv\mathcal{T}_a$ is used for the cosmic clock, since this is independent of the choice of non-linear definition.

\section{Explicit results in conformal-Newtonian gauge}
\label{sec:cNresult}

To get insights into the gauge-invariant ruler quantities $\mathcal{C}$, $\mathcal{M}$ and $\mathcal{T}$,
we find it instructive to work out the observed galaxy density $\dgobs$ in the conformal Newtonian (cN) gauge. Consider a perturbed FLRW metric written in the general form
\begin{equation}
    \label{eq:perturbed metric}
    \mathrm{d}s^2 = a^2(\eta)\Big[-(1+2A)\mathrm{d}\eta^2 - 2B_i\mathrm{d}\eta\mathrm{d}x^i + (\delta_{ij} + h_{ij})\mathrm{d}x^i\mathrm{d}x^j\Big] \;,
\end{equation}
where $\eta$ denotes conformal time and $a(\eta)$ is the scale factor. The cN gauge is obtained upon setting $A=\Psi$, $h_{ij}=-2\Phi\delta_{ij}$ and all the other scalar degrees of freedom to zero. Here, we only consider scalar perturbations in $B_i$ and $h_{ij}$.

Furthermore, since the observer naturally measures galaxy number counts {\it per observed redshift range and observed solid angle on the sky}, it is desirable to use the local coordinates $\tilde x^i=(\tilde z,\ntvh)$ defined at the observer. The corresponding over-density $\dgobs(\tilde z,\ntvh)$ can be obtained by following the procedure outlined in Section \ref{sec:gauge invariant expression}, that is, by pulling back $(k\cdot u_g)\omtau$ using the local coordinates $(\tilde{z},\ntvh)$. Alternatively, one can also start from the Jacobi map expression Eq.~\eqref{eq:dNdzdO},
\begin{equation}
    \frac{\mathrm{d}N}{\mathrm{d}\tilde{z}\mathrm{d}\tilde \Omega}= n_g \bigg\lvert\frac{\mathrm{d}\ell}{\mathrm{d}\tilde z}\bigg\lvert \det\!\mathcal D_o\;,
\end{equation}
and successively compute the line-of-sight (parallel) and transverse parts $n_g\big\lvert\frac{\mathrm{d}\ell}{\mathrm{d}\tilde z}\big\lvert$ and $\det\!\mathcal D_o$, respectively. This was the approach taken by Challinor and Lewis \cite{ChallinorLewis2011}. The two approaches agree with each other precisely because $(k\cdot u_g)\omtau$ naturally splits as $\mathrm{i}\,\mathrm{d}\ell \wedge \mathrm{d}\zeta \wedge \mathrm{d}\bar{\zeta}$.

Here, we shall revisit the calculation of ref. \cite{ChallinorLewis2011} in order to emphasise the interplay between the rulers, the distortions of the geodesic congruence, the Hubble parameter, and the angular-diameter distance defined in the observer's reference frame. Since the latter is naturally endowed with a set of conformal Fermi coordinates (CFC), we shall refer to it as the CFC frame \cite{Dai:2015jaa}. Furthermore, we shall work to linear order in perturbations and on a spatially flat background FLRW universe throughout the rest of the paper.

\subsection{Line-of-sight perturbations and local Hubble parameter}

We start with the computation of $n_g\big\lvert\frac{\mathrm{d}\ell}{\mathrm{d}\tilde z}\big\lvert$.
For shorthand convenience, let $a_s\equiv a(\eta(\tau_s))$ and $a_o\equiv a(\eta(\tau_o))$ be the scale factor at the time of emission and measurement as defined by the source and observer proper time $\tau_s$ and $\tau_o$, respectively (these are \emph{not} equal to $\at(\tau_s)$ and $\at(\tau_o)$, in general). When solving the geodesic equation, one finds that the observed redshift $\tilde z$ is given by
\begin{equation}
    \label{eq:redshift}
    1+\tilde z = \frac{a_o}{a_s}\bigg\{1+\Psi_o-\Psi_s-\int_0^{\chi_s}\!\mathrm{d}\chi'\,\big(\dot{\Psi}+\dot{\Phi}\big) + v_{\parallel s} - v_{\parallel o}\bigg\}\;,
\end{equation}
where a dot denotes a derivative with respect to the conformal time coordinate.
Here and henceforth, subscript ``s" or ``o" designate quantities evaluated at the source and observer position, respectively.
In linear theory, we can set $\chi_s\equiv \tilde\chi$ where $\tilde\chi=\chi(\tilde z)$ is the inferred comoving distance to redshift $\tilde z$.
Furthermore, $v_{\parallel}$ is the line-of-sight peculiar velocity and $\ntvh\equiv\nvh$ is the sky direction as seen by the observer (the photon's direction of propagation is $-\nvh$ in the first-order Born approximation).

Eq.~\eqref{eq:redshift} can be used to define the perturbation $\delta\eta=\delta\eta(\tilde z,\tilde\chi\ntvh)$ to the time coordinate on hyper-surfaces of constant observed redshift $\tilde z$. To proceed, it is essential to enforce that the observer be on a hyper-surface of zero observed redshift, i.e. $\tilde z \equiv 0$. On writing $\eta(\tau_s) \equiv \tilde\eta + \delta\eta$ where $\tilde\eta\equiv\bar\eta(\tilde z)$ is
the conformal time coordinate corresponding to the redshift $\tilde z$ in the unperturbed background, the source scale factor is
\begin{equation}
\label{eq:as}
    a_s = a(\tilde\eta + \delta\eta)\approx \tilde a \big(1+\tilde{\cal H} \delta\eta\big)\;,
\end{equation}
where ${\cal H} = \dot{a}(\bar\eta)/a(\bar\eta)$ is the conformal FLRW Hubble rate.
A tilde will denote {\it background} quantities evaluated at conformal time $\tilde\eta$, so that $\tilde a \equiv a(\tilde\eta)$, \emph{etc.} On writing the time coordinate of the photon's geodesic at the observer as $\eta(\tau_o)=\bar\eta_o+\delta\eta_o$, a similar calculation at the observer position yields
\begin{equation}
\label{eq:ao}
    a_o = a(\bar{\eta}_o + \delta\eta_o) \approx 1+ {\cal H}_o\delta\eta_o \;,
\end{equation}
where ${\cal H}_o = {\cal H}(\bar\eta_o)$ is the present-day FLRW expansion rate and $\delta\eta_o$ is $\delta\eta$ evaluated at the observer's position. Note that the observer defined by $\tilde{z}=0$ has none-zero time coordinate perturbation $\delta\eta_o$ compared to the background relation $\bar{\eta}(z=0)$.

The perturbation $\delta\eta_o$ to the observer's time coordinate on the hyper-surface $\tilde z \equiv 0$ can be determined through the requirement that the measurement be performed at fixed observer's proper time (which corresponds to the condition $V=$~const., see Section \ref{sec:invariant volume forms}), that is,
\begin{equation}
    \delta\left(\int_0^{\tau_o}\!\mathrm{d}\tau\right)=\delta\left(\int_0^{\eta_o}\!\mathrm{d}\eta\, \frac{\sqrt{-g_{\mu\nu}u^\mu u^\nu}}{u^0}\right)\equiv 0 \;,
\end{equation}
At first order in perturbations, this implies
\begin{equation}
    \delta\eta_o = - \int_0^{\bar\eta_o}\!\mathrm{d}\eta\,a(\eta)\,\Psi({\bf 0},\eta)
\end{equation}
for an observer located at the origin of the coordinate system at the time of the measurement.

Substituting these expressions into Eq.~\eqref{eq:redshift} and taking advantage of the fact that
\begin{equation}
    1+\tilde z \equiv \frac{1}{\tilde a}\;,
\end{equation}
the perturbation $\delta\eta$ to the source time coordinate is given by
\begin{equation}
    \label{eq:deta}
    \tilde{\cal H}\delta\eta = \Psi_o-\Psi-\int_0^{\tilde\chi}\!d\chi\,\big(\dot{\Psi}+\dot{\Phi}\big) + v_\parallel - v_{\parallel o}+{\cal H}_o\delta\eta_o \;,
\end{equation}
where $\tilde\chi\equiv \bar\eta_o-\tilde\eta$ is  the  radial  comoving  coordinate  of  the  constant observed redshift hyper-surface in the background spacetime\footnote{We have $\tilde\eta\equiv \eta_*$ and $\tilde\chi\equiv\chi_*$ in the notation of \cite{ChallinorLewis2011}.} and, hereafter, we omit the subscript `s' to avoid clutter.
Furthermore, except for the integrals along photon geodesics, all the source and observer terms are evaluated at the spacetime coordinates $(\tilde\eta,\tilde\chi\ntvh)$ and $(\bar\eta_o,{\bf 0})$, respectively.
While Eq.~\eqref{eq:deta} differs from the expression given in \cite{ChallinorLewis2011} owing to the last contribution, it precisely agrees with the variable $(\Delta\ln a)_\text{cN}$ introduced in \cite{SchmidtJeong2012} and in appendix \ref{sec:alternative}. The presence of ${\cal H}_o\delta\eta_o$ is crucial to ensure the gauge-invariance of $\dgobs$ \citep[see, e.g.][for early discussions]{SchmidtJeong2012,Biern:2016kys}.

At this point, we can take the derivative of \eqref{eq:redshift} with respect to the affine parameter of the light ray and calculate $\frac{\mathrm{d}\tilde z}{\mathrm{d}\lambda}$ at fixed observed redshift. On writing $k^0 = \frac{\mathrm{d}\eta}{\mathrm{d}\lambda}=\frac{E_\nu}{a}(1-\Psi)$ where $E_\nu$ is the (physical) photon energy measured by a local comoving observer, we find
\begin{equation}
  \label{eq:dzdlambda}
  \frac{\mathrm{d}\tilde z}{\mathrm{d}\lambda} = - \frac{\tilde{\cal H} E_\nu}{\tilde a^2}\bigg\{1+ \left(\frac{\dot{\tilde{\cal H}}}{\tilde{\cal H}}-\tilde{\cal H}\right)\delta\eta + \frac{1}{\cal H}\bigg[\frac{\mathrm{d}\Psi}{\mathrm{d}\eta}-\dot\Psi-\dot\Phi-\frac{\mathrm{d}v_\parallel}{\mathrm{d}\eta}\bigg]-\Psi\bigg\} \;.
\end{equation}
Here, the directional derivative $\frac{\mathrm{d}X}{\mathrm{d}\eta}=\dot X - (\ntvh\cdot\nabla) X$
is computed along incoming photon geodesics, and all the perturbations are evaluated at coordinate $(\tilde\eta,\tilde\chi\nvh)$.
Using ${\rm d}\ell = - k_\mu u_g^\mu \mathrm{d}\lambda$ (where $\mathrm{d}\ell>0$ in the propagation direction of the light beam) and $k_\mu u_g^\mu = -E_\nu(1+v_\parallel)$, we arrive at
\begin{equation}
\label{eq:nxdldz1}
    n_g(\tvx)\bigg\lvert\frac{\mathrm{d}\ell}{\mathrm{d}\tilde z}\bigg\lvert = \frac{\tilde a^2\bar n_g(\tilde z)}{\tilde{\cal H}}
    \bigg\{1+\delta_g^\textrm{or}-\left(\frac{\dot{\tilde{\cal H}}}{\tilde{\cal H}}-\tilde{\cal H}\right)\delta\eta-\frac{1}{\tilde{\cal H}}\frac{\mathrm{d}\Psi}{\mathrm{d}\eta}
    +\frac{1}{\tilde{\cal H}}\big(\dot{\Psi}+\dot{\Phi}\big)+\frac{1}{\tilde{\cal H}}\frac{\mathrm{d}v_\parallel}{\mathrm{d}\eta}+\Psi+v_\parallel\bigg\}
\end{equation}
The gauge-invariant perturbation $\delta_g^\textrm{or}$ (density contrast defined in the constant-observed-redshift gauge) is related to the synchronous gauge matter perturbation $\delta_m^\textrm{syn}$ at constant coordinate time through
\begin{equation}\label{eqn: equation 4.12}
    \delta_g^\text{or} = b_1\delta_m^\textrm{syn} + \frac{\mathrm{d}\ln {\bar n}_g}{\mathrm{d}\ln \tilde{a}}\mathcal{T} \;,
\end{equation}
where $b_1$ is the linear galaxy bias.

To understand how $n_g(\tvx)\left|\frac{\mathrm{d}\ell}{\mathrm{d}\tilde z}\right|$ behaves in the limit $\tilde z\to 0$, consider a source at affine parameter $\lambda = \lambda_o - \mathrm{d}\lambda$, where $\lambda_o$ is the value at the observer position (the focus point of the null congruence). At first order in $\mathrm{d}\lambda$, \eqref{eq:dzdlambda} yields
\begin{equation}
    \label{eq:localdzdl1}
    \mathrm{d}\tilde z = {\cal H}_o\left\{1+\left(\frac{\dot{\cal H}_o}{{\cal H}_o}-{\cal H}_o\right)\delta\eta_o-\Psi_o-\frac{1}{{\cal H}_o} \Big[\dot\Phi_o+\big(\partial_\parallel\Psi\big)_o+\dot v_{\parallel o}-\big(\partial_\parallel v_\parallel\big)_o\Big]-v_{\parallel o}\right\}\, |\mathrm{d}\ell|\;,
\end{equation}
where $\partial_\parallel =\ntvh\cdot\grad$ is taken along the line-of-sight.
The monopole of this local distance-redshift relation precisely corresponds to the CFC Hubble parameter of Ref. \cite{Dai:2015jaa}
\begin{align}
\label{eq:HF}
\HF &\equiv \nabla_\mu u_o^\mu (\tau_o) \\
&= {\cal H}_o\bigg[1+\left(\frac{\dot{\cal H}_o}{{\cal H}_o}-{\cal H}_o\right)\delta\eta_o- \Psi_o - \frac{1}{{\cal H}_o}\dot\Phi_o + \frac{1}{3{\cal H}_o} \big(\partial_i v^i\big)_o\bigg] \nonumber
\end{align}
Therefore, \eqref{eq:localdzdl1} can be recast into the form
\begin{equation}
    \label{eq:localdzdl2}
    \mathrm{d}\tilde z = \Big(\HF + \mbox{dipole}(\hat{\bm{n}}) + \mbox{quadrupole}(\hat{\bm{n}}) \Big)\, |\mathrm{d}\ell|\;.
\end{equation}
This emphasises that the local, sky-average expansion rate, inferred from local distance and redshift measurements, is, fundamentally, the CFC Hubble parameter $\HF$.

Based on these findings, we can split the observable quantity $n_g(\tvx)\left|\frac{\mathrm{d}\ell}{\mathrm{d}\tilde z}\right|$ into an average and a (gauge-invariant) fluctuation in two different ways.\footnote{We prove explicitly that $\mathcal{C}$ is gauge-invariant in any gauge in appendix \ref{sec:C gauge invariance}.}
On expressing $\mathcal{C}$ and $\mathcal{T}$ in conformal Newtonian gauge,
\begin{align}
\label{eq:CrulercN}
    \big(\mathcal{C}-\mathcal{T}\big)_\text{cN} &\equiv
    \left(\frac{\dot{\tilde{\cal H}}}{\tilde{\cal H}}-\tilde{\cal H}\right)\delta\eta
    +\frac{1}{\tilde{\cal H}}\frac{d\Psi}{d\eta} -\frac{1}{\tilde{\cal H}}\big(\dot{\Psi}+\dot{\Phi}\big)-\Psi
-\frac{1}{\tilde{\cal H}}\frac{\mathrm{d}v_\parallel}{\mathrm{d}\eta}-v_\parallel \;,  \\
\label{eq:Co}
\big(\mathcal{C}_o\big)_\text{cN} &=\left(\frac{\dot{\cal H}_o}{{\cal H}_o}-{\cal H}_o\right)\delta\eta_o- \Psi_o - \frac{1}{{\cal H}_o}\dot\Phi_o + \frac{1}{3{\cal H}_o} \big(\partial_i v^i\big)_o\;,
\end{align}
and using the fact that $\tilde{\cal H}/{\cal H}_o = E(\tilde z)/(1+\tilde z)$, where
\begin{equation}
    E(\tilde z) = \sqrt{\Omega_m (1+\tilde z)^3 + \Omega_\Lambda + \Omega_{\rm rad}(1+\tilde{z})^4 + \Omega_K(1+\tilde{z})^2},
\end{equation}
we eventually obtain (we omit the subscript ``cN'' since this decomposition truly is gauge-invariant)
\begin{equation}
\label{eq:nxdldz2}
\begin{aligned}
\mbox{FLRW}:\quad
    n_g(\tvx)\bigg\lvert\frac{\mathrm{d}\ell}{\mathrm{d}\tilde z}\bigg\lvert &= \frac{(1+\tilde z)\bar n_g(\tilde z)}{\mathcal{H}_o E(\tilde z)}\Big\{1-\mathcal{C}
    +\mathcal{T}+ \delta_g^\textrm{or}\Big\} \\
\mbox{CFC}:\quad
    n_g(\tvx)\bigg\lvert\frac{\mathrm{d}\ell}{\mathrm{d}\tilde z}\bigg\lvert &= \frac{(1+\tilde z)\bar n_g(\tilde z)}{\HF E(\tilde z)}\Big\{1-\big(\mathcal{C}-\mathcal{C}_o\big)
    +\mathcal{T}+\delta_g^\textrm{or}\Big\}     \;.
\end{aligned}
\end{equation}
In Eq.~\eqref{eq:nxdldz2}, both choices of background (labelled as ``FLRW'' or ``CFC'') are admissible as $\tilde z$ is the observed redshift, the average galaxy density $\bar n_g$ can generally be predicted from a first-principle galaxy formation theory, the function $E(\tilde z)$ depends only on background quantities such as $\Omega_m$, $\Omega_\Lambda$ \emph{etc.} and, in the ``CFC'' case, $\HF$ is the local Hubble parameter directly measured by the observer. In the latter case, the local value of the monopole of $\mathcal{C}$ is absorbed into the local Hubble rate.

The curly brackets of Eq.~\eqref{eq:nxdldz2} define the line-of-sight contribution to the observed galaxy over-density. For the FLRW background, this over-density differs somewhat from the expression quoted in \cite{ChallinorLewis2011} (which is the curly bracket of Eq.~\eqref{eq:nxdldz1}) owing to an additional perturbation ${\cal H}_o\delta\eta_o$ arising from the requirement that the measurement be performed at fixed $\tau_o$.

\subsection{Transverse perturbations and local angular-diameter distance}

We now turn to the transverse part, $\det\!\mathcal{D}_o$.
At linear order, the determinant $\det\!\mathcal{D}_o(\tilde z,\nvh)$ of the Jacobi map in the local coordinates $\tilde x^i=(\tilde z,\ntvh)$ simplifies to \cite{ChallinorLewis2011}
\begin{equation}
    \label{eq:detjacobimap1}
    \det\!\mathcal{D}_o(\tilde z,\chi\nvh)= \tilde a^2\tilde\chi^2\left(1+2\frac{\delta\chi}{\tilde\chi}+2\tilde{\cal H}\delta\eta - 2\Phi -2\kappa +2 v_{\parallel o}\right) \;,
\end{equation}
where
\begin{equation}
    \label{eq:kappa}
    \kappa(\eta,\chi\nvh) = \frac{1}{2\tilde\chi}\int_0^{\tilde\chi}\!\mathrm{d}\chi' \chi'\big(\chi-\chi'\big)\partial_\perp^2\big(\Psi+\Phi\big)
\end{equation}
is the lensing convergence and
\begin{equation}
    \label{eq:dchi}
    \delta\chi = \delta\eta_o -\delta\eta + \int_0^{\tilde\chi}\!\mathrm{d}\chi\,\big(\Psi+\Phi\big)
\end{equation}
is the perturbation to the radial coordinate (at constant observed redshift), which straightforwardly follows from the radial part of the perturbed metric in the Born approximation. Note the additional $+ \,\delta\eta_o$ relative to the expression given in \cite{ChallinorLewis2011}.

As emphasised in Section \ref{sec:invariant volume forms}, $\sqrt{\det\!\mathcal D_o}$ is a perturbed version of the angular diameter distance. In the limit of small redshift $\tilde z\ll 1$, it reduces to the physical distance $\ell$ measured along the light beam. Since $\HF$ is the expansion rate locally measured by the observer, we must have
\begin{equation}
    \sqrt{\det\!\mathcal D_o}\approx \frac{\tilde z}{\HF} \quad \mbox{for}\quad \tilde z\ll 1\;.
\end{equation}
Therefore, we introduce the CFC angular-diameter distance
\begin{align}
    \label{eq:DAF}
    \DAF(\tilde z) \equiv \frac{{\cal H}_o}{\HF}\tilde a \tilde\chi =\frac{1}{(1+\tilde z)\HF}\int_0^{\tilde z}\!\mathrm{d}z\, \frac{1+z}{E(z)} \;,
\end{align}
which differs from the FLRW angular-diameter distance $d_A(\tilde z)$ owing to the presence of $\HF$. Here, we're using $E(z)$ from a previously known background. A third alternative would be to perform a genuine measurement of $E(z)$ using the monopole of the luminosity distance at each redshift, in which case $\mathcal{M}$ would have a vanishing monopole at each redshift. Besides, in the limit $\tilde\chi=\bar\eta_o-\tilde\eta \ll 1$ of short separation between the source and observer, the determinant of the Jacobi map reduces to
\begin{align}
    \label{eq:detjacobimap2}
    \sqrt{\det\!\mathcal D_o} &\approx \tilde\chi \bigg\{1+{\cal H}_o\delta\eta_o+\left(\Psi_o+\frac{1}{{\cal H}_o} \dot{\Phi}_o\right)+\frac{1}{{\cal H}_o}\Big[\big(\partial_\parallel\Psi\big)_o+\dot{v}_{\parallel o} - \big(\partial_\parallel v_\parallel\big)_o\Big]+v_{\parallel o} \bigg\}
\end{align}
at first order in $\tilde\chi$ since the lensing convergence scales like $\kappa\propto\tilde\chi^2$ in that limit. This relation emerges from
\begin{align}
    \left(\frac{\delta\chi}{\tilde\chi}- \Phi\right)\bigg\lvert_o &\approx \frac{1}{\mathcal H_o\tilde\chi}\Big[\dot{\Phi}_o+\big(\partial_\parallel\Psi\big)_o-\big(\partial_\parallel v\big)_o+\dot{v}_{\parallel o}\Big]\big(\bar\eta_o-\tilde\eta\big)+\frac{1}{\tilde\chi}\big(\Psi_o+\Phi_o\big)\big(\bar\eta_o-\tilde\eta\big)-\Phi_o \nonumber \\
    &= \Psi_o + \frac{1}{\mathcal H_o}\dot{\Phi}_o+\frac{1}{\mathcal H_o}\Big[\big(\partial_\parallel\Psi\big)_o-\big(\partial_\parallel v_\parallel\big)_o + \dot{v}_{\parallel o}\Big].
\end{align}
Taking into account that the observed redshift is $\tilde z\approx \frac{\dot a_o}{a_o}\tilde\chi$, that is,
\begin{equation}
    \tilde\chi \approx \frac{\tilde z}{{\cal H}_o}\left(1-\frac{\dot{\cal H}_o}{{\cal H}_o}\delta\eta_o\right)\;,
\end{equation}
we see that, in the limit $\tilde z\to 0$, the monopole of Eq.~\eqref{eq:detjacobimap2} exactly returns $\DAF(\tilde z)$.

Therefore, on expressing $\mathcal{M}$ in conformal Newtonian gauge \cite{SchmidtJeong2012},
\begin{align}
\label{eq:MrulercN}
    \big(\mathcal{M}-2\mathcal{T}\big)_\text{cN} &\equiv-2\frac{\delta\chi}{\tilde\chi}-2\tilde{\cal H}\delta\eta +2 \Phi+2\big(-v_{\parallel o}+\kappa\big)  \\
    \label{eq:Mo}
    \big(\mathcal{M}_o\big)_\text{cN} &=2 \big(\mathcal{C}_o\big)_\text{cN} \;,
\end{align}
the transverse contribution to the galaxy count can be expressed as
\begin{equation}
\label{eq:detjacobimap3}
\begin{aligned}
    \mbox{FLRW}:\quad \det\!\mathcal{D}_o(\tilde z,\nvh) &= \big[d_A(\tilde z)\big]^2\Big\{1 - \mathcal{M}+2
    \mathcal{T}\Big\} \\
    \mbox{CFC}:\quad \det\!\mathcal{D}_o(\tilde z,\nvh) &= \big[\DAF(\tilde z)\big]^2\Big\{1-\big(\mathcal{M}-\mathcal{M}_o\big)+2
    \mathcal{T}\Big\}
\end{aligned}
\end{equation}
upon substituting either the FLRW or the CFC angular-diameter distance into Eq.~\eqref{eq:detjacobimap1}. Here again, both choices are perfectly admissible.
Note that the convergence $\kappa$ scales as $\tilde z^2$ in the limit $\tilde z\ll 1$ whereas the first three terms in the curly brackets of Eq.~\eqref{eq:MrulercN} vanish linearly in $\tilde z$.

\subsection{Choice of background and rulers}
\label{sec:delta obs with cN rulers}

Our cN gauge calculation has elucidated the dependence of the observed galaxy density $\tilde n_g$ on ``background'' quantities. We have shown that the observed galaxy density {\it per unit redshift and observed solid angle on the sky} can be cast into the form
\begin{align}
\label{eq:finaldeltagrulers}
 \mbox{FLRW}:\quad \tilde n_g(\tilde z,\ntvh) &= \bar n_g(\tilde z) \frac{(1+\tilde z)\big[d_A(\tilde z)\big]^2}{E(\tilde z)\,\mathcal{H}_o}\,\Big\{1 - \mathcal{C} - \mathcal{M} + 3\mathcal{T} + \delta_g^\textrm{or}\Big\}  \\
  \mbox{CFC}:\quad \tilde n_g(\tilde z,\ntvh) &= \bar n_g(\tilde z) \frac{(1+\tilde z)\big[\DAF(\tilde z)\big]^2}{E(\tilde z)\,\HF}\,\Big\{1 - \big(\mathcal{C} -\mathcal{C}_o\big)- \big(\mathcal{M} -\mathcal{M}_o\big)+
  3\mathcal{T} + \delta_g^\textrm{or}\Big\} \nonumber
\end{align}
where the overall multiplicative factor involves, in addition to the average galaxy density $\bar n_g$ and the scaled Hubble rate $E(\tilde z)$, either the FLRW or the CFC Hubble parameter and angular-diameter distance. Note that both $\HF$ and $\DAF(\tilde z)$ are defined through a local measurement of the expansion rate and, therefore, are gauge-independent. This is not at odds with the fact that they depend on observer terms explicitly. When the CFC background is selected, the appearance of $\HF$ and $\DAF(\tilde z)$ ensures that the monopole of the parallel and transverse projection terms vanishes in the limit $\tilde z\to 0$.
Both choices of background are perfectly admissible. They are nearly equivalent in the particular case of an EdS cosmology, for which $\mathcal{M}_o=2\mathcal{C}_o=\frac{2}{3\mathcal{H}_o}\partial_i v_o^i$ (see Appendix \ref{sec:rulers in EdS}). Therefore, the definition of the rulers relies on the choice of background: a God-given true (global) background or the (CFC) one defined by local measurements.
Finally, for sake of completeness, $\big(\mathcal{T}\big)_\text{cN}$ is given by \cite[see equation 15 of][]{JeongSchmidt2014}
\begin{equation}
\label{eq:TrulercN}
\begin{aligned}
        \big(\mathcal{T}\big)_\text{cN} &\equiv \frac{\tilde{\cal H}}{\tilde a}\int_0^{\tilde\eta}\!\mathrm{d}\eta'\,\Psi(\tilde\vx,\eta')a(\eta')- {\mathcal H}_o\int_0^{\bar\eta_o}\!\mathrm{d}\eta'\,\Psi({\bf 0},\eta') a(\eta') \\
        &\qquad +\Psi_o-\Psi+v_\parallel - v_{\parallel o}+\int_0^{\tilde\eta}\!\mathrm{d}\eta'\, \big(\dot\Psi+\dot\Phi\big) \;.
\end{aligned}
\end{equation}
Note that $\big(\mathcal{T}\big)_\text{cN}\to 0$ in the limit $\tilde z\to 0$.

The invariance of $\mathcal{C}$, $\mathcal{M}$ and $\mathcal{T}$ under the addition of a constant + pure gradient metric perturbations (as demonstrated in \cite{SchmidtJeong2012} for the particular case of an EdS cosmology) implies, \emph{inter alia}, that the observed galaxy counts $\tilde n_g(\tilde z,\ntvh)$ are invariant under a shift of the Bardeen potentials $\Psi\to\Psi+\Psi_\ell$ and $\Phi\to\Phi+\Phi_\ell$ \cite[see the discussion in, e.g.,][]{Peloso:2013zw,Kehagias:2013yd,Creminelli:2013mca}. Such a transformation amounts to a re-parameterisation of the proper time and length. Therefore, it affects neither $\HF$ nor $\DAF(\tilde z)$. This invariance meshes well with the intuition that a uniform potential does not have any physical impact on the observed galaxy distribution.

\section{The low multipoles of the observed angular power spectra}
\label{sec:observed low multipoles}

In this section, we present the low multipoles of the observed galaxy power spectrum. For simplicity, we specialise to the case of an EdS cosmology. Details of the calculation can be found in Appendix \ref{sec:rulers in EdS}.

\begin{figure}
    \centering
    \includegraphics[width=0.7\textwidth]{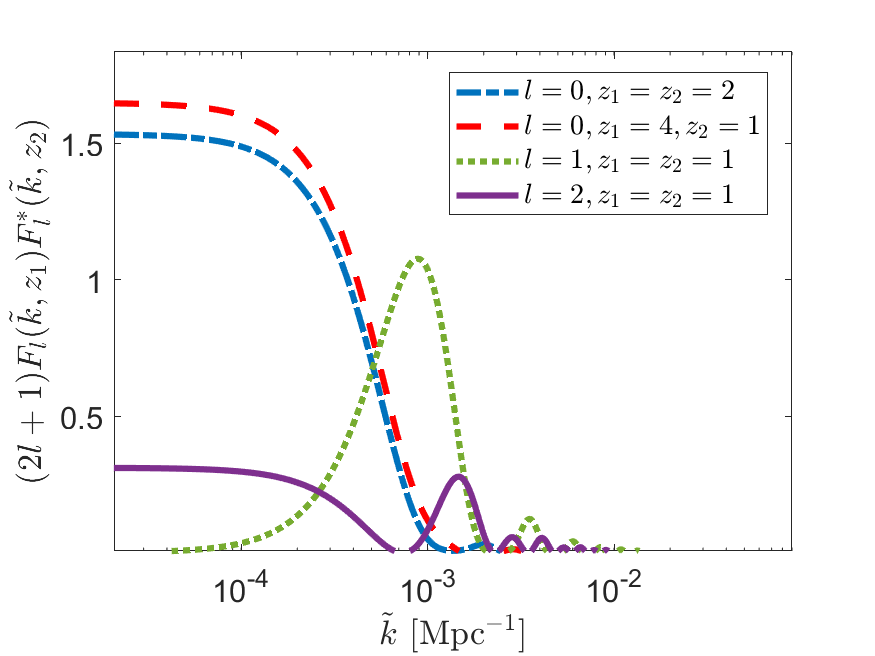}
    \caption{The integrand defining the monopole, dipole and quadrupole for various redshifts up to an overall normalisation, for the FLRW case in equation \eqref{eq:finaldeltagrulers}.  Results are shown for EdS cosmology.}
    \label{fig:c_ells}
\end{figure}

As $\dgobs(\tilde z,\ntvh)$ is an observable, its angular power-spectrum $C_l$ must be well-defined for all multipoles $l$, including $l = 0$. On writing
\begin{equation}
    \dgobs(\tilde{z},\ntvh) = \sum_{l m}\big(\dgobs\big)_{l m}\!(\tilde{z})\, Y_{\ell m}(\ntvh) \;,
\end{equation}
the angular correlator defined by
\begin{equation}
    \big\langle \big(\dgobs\big)_{l m}(z_1)\big(\dgobs\big)^*_{l' m'}(z_2)\big\rangle = C_l(z_1,z_2) \delta_{l l'}\delta_{mm'}
\end{equation}
becomes
\begin{equation}
    \label{eq:C_l}
    C_l(z_1,z_2) = \frac{2}{\pi}\int_0^\infty\!\!\mathrm{d}k\, k^2 F_l(k\chi_1)F_l(k\chi_2)\, P_{m,o}(k)\;,
\end{equation}
where $P_{m,o}(k)$ is the present-day matter power spectrum in synchronous comoving gauge.
The real-valued function $F_l(k\chi)$ is a combination of spherical Bessel functions, their derivatives and integrals, \emph{etc.}, and may be worked out from the explicit expressions given in Appendix \ref{sec:rulers in EdS}.

In figure \ref{fig:c_ells} we show the integrand defining the monopole $C_{l=0}$ as well as the dipole and quadrupole $C_{l=1}$ and $C_{l=2}$ for comparison, for a few choices of redshifts. We assume an EdS cosmology for which $\Psi=\Phi$ is time-independent. The invariance of the observed $C_l$ under a shift of the gravitational potential $\Psi$ by a constant or a pure gradient term regularises the low-$k$ limit of $F_0$ (through the dependence on $\Psi-\Psi_o$) and of $F_1$, so that, unlike Refs.~\cite{Bertacca:2012tp,Raccanelli:2013dza,Bertacca:2019wyg} for instance, there is no need for an infrared cut-off.
Therefore, our $C_{l=0}$ indeed represents the variance of the observed monopole as measured by different observers with same proper time $\tau_o$.
For the dipole, the integrand in the limit $k\to 0$ vanishes because $F_1$ is odd under a parity transformation whereas, for the quadrupole, $F_2$ asymptotes to a constant as $C_{l=2}$ arises from density and tidal shear fluctuations.

\section{Discussion and Conclusion}
\label{sec:discussion}

In this paper, we have revisited the signature of projection effects on the observed galaxy over-density $\dgobs$ in order to establish a clear connection between different approaches and clarify the role of observer terms. In particular, we have presented
\begin{itemize}
    \item A derivation of the observed galaxy counts and galaxy over-density using the Newman-Penrose formalism \cite{Newman:1961qr} that exemplifies the relation between the Jacobi map and the cosmic-ruler approach of refs. \cite{ChallinorLewis2011} and \cite{Jeong:2011as}, respectively.
    \item An explicit expression for the \emph{non-linear} galaxy over-density in terms of the gauge-invariant ruler quantities introduced in refs. \cite{SchmidtJeong2012,JeongSchmidt2014} that provides a starting point for higher-order calculations.
    \item A linear theory computation of the observed galaxy over-density in the conformal Newtonian gauge which elucidates the relation between the rulers and the local CFC background defined by the observer.
\end{itemize}
Our non-linear expression \eqref{eq:over density non linear} is valid at any order in perturbations. Therefore, it provides a convenient starting point for a computation of the non-linear galaxy number counts in terms of the gauge-invariant rulers $\mathcal{C}$, $\mathcal{B}_i$, $\mathcal{A}_{ij}$, and $\mathcal{T}_a$. This should be especially useful for the inclusion of  projection effects in the observed galaxy bispectrum \cite[see, e.g.,][for existing calculations]{Bertacca:2014dra,DiDio:2014lka,DiDio:2015bua,Kehagias:2015tda,Umeh:2016nuh} and higher correlation functions. Such a calculation would proceed as follows: first, compute $\mathcal{C}$, $\mathcal{T}$, $\mathcal{B}_i$ and $\mathcal{A}_{ij}$ using equations \eqref{eq:ruler definitions} and \eqref{eq:defTa} to the appropriate order in the cosmological perturbations, by solving the geodesic equation; the gauge-invariance of these rulers/clocks allows one to work in any gauge. For example, for $\mathcal{T}$, one only needs the proper time perturbation. Then, to find $\dgobs$, all one needs to do is to substitute in equation  \eqref{eq:over density non linear}, and keep terms up to the desired order.

The galaxy over-density computed here explicitly depends on quantities evaluated at the observer's position. As a result, the ensemble corresponding to statistics averaged over survey volumes is a constrained ensemble as outlined in \cite{Desjacques:2020zue}.
This originates from the existence of an observer's viewpoint that constrains the fluctuation fields $f(\eta,\vx)$ which $\dgobs$ explicitly depends on to assume specific values $f_o$ at the (spacetime) observer's position $x^\mu(\tau_o)\equiv x_o^\mu$.
If $f$ is locally observable---like the peculiar velocity $\vv$ (relative to the cosmic microwave background or CMB) frame for instance---then the value $f_o\equiv f(x_o^\mu)\ne 0$ can be determined from the nearby matter distribution.
If $f$ is not locally observable--- as is the case of the Bardeen potentials $\Psi$ and $\Phi$---then differences of the form $f\equiv \Psi-\Psi_o$, solely, are measurable, whence $f_o\equiv 0$. The observer's constraint thus is compatible with the invariance of $\dgobs$ under an arbitrary constant or pure gradient shift of the potentials.
At the level of the monopole $C_{l=0}$ of the galaxy over-density field, this invariance ensures that the Fourier-space kernel (proportional to $F_0(k\chi_1)F_0(k\chi_2)$, see Section \ref{sec:observed low multipoles}) does not diverge in the limit $k\to 0$. This also guarantees that the local background ``seen'' by the observer matches the local CFC frame \citep[see][]{Dai:2015jaa}.

Furthermore, we have shown that the local Hubble parameter $\HF=\nabla_\mu u^\mu(\tau_o)$ introduced in \cite{Dai:2015jaa} is the quantity fundamentally extracted from the observed-distance-redshift relations, in agreement with ref. \cite{Heinesen:2020bej} \citep[see also][for related discussions]{Lombriser:2019ahl}. As a perfectly local measurement, $\HF$ depends on the local divergence $(\partial_iv^i)_o$ of the galaxy peculiar velocity field, which has a short coherence length (relative to the metric perturbations). As a result, $(\partial_iv^i)_o$ is dominated by non-linear small-scale fluctuations.
In practice, one attempts to correct the redshifts of local standard candles
for their peculiar velocities based on observations of local structures \cite[e.g.][]{cooray/caldwell:2006,davis/hui/etal:2011,marra/etal:2013,riess/etal:2016}. It
is highly non-trivial to identify a covariant relativistic observable that
corresponds to this, now non-local, measurement of the Hubble parameter, since that involves
constructing a relativistic embedding of the inference of the velocity field
from the observed galaxy density, which is used to correct redshifts for
peculiar velocities.

Let us instead give an educated guess for what the result will approximately
be, in the context of linear perturbations in cN gauge.
This guess consists of simply dropping the local velocity divergence from the CFC Hubble rate at the observer:
\begin{equation}\label{eq:reconstrcuted Hubble rate}
    \frac{\dot a_o}{a_o}\left(1-\Psi_o-\frac{1}{{\cal H}_o}\dot\Phi_o\right).
\end{equation}
This expression is not generally covariant, but we expect it to be fairly
accurate as long as all small-scale perturbations are removed,
and non-linear effects can be neglected. To understand this expression,
consider an observer who uses a local FLRW metric $\mathrm{d}s^2 = -dt_L^2 + a_L^2(t_L)\, \mathrm{d}\vx_L^2$ to measure time and length intervals in a small region of spacetime centred on her position $x_o^\mu$. She would indeed measure the local Hubble parameter as
\begin{equation}
        \frac{1}{a_L}\frac{\mathrm{d}a_L}{\mathrm{d}t_L}
        \approx \frac{\big(1+\Phi_o\big)}{a_o}  \frac{\mathrm{d}}{\mathrm{d}\eta}\Big[a_o\big(1-\Phi_o\big)\Big]  \frac{\mathrm{d}\eta}{\mathrm{d}t_L}
\end{equation}
at linear order. For a standard $\Lambda$CDM cosmology with scalar metric fluctuations of order $10^{-5} - 10^{-4}$, equation \eqref{eq:reconstrcuted Hubble rate} differs only slightly from the FLRW value.
Note that large-scale potential perturbations whose wavelength is much longer than the largest distances between source and observer do not contribute to equation \eqref{eq:reconstrcuted Hubble rate}, just like in the case of the local CFC Hubble rate. Since small-scale perturbations are removed by the idealised peculiar velocity correction, equation \eqref{eq:reconstrcuted Hubble rate} differs from the underlying background value only by the contributions from an intermediate range of wave numbers.


\acknowledgments

We are grateful to Lea Reuveni for making figure \ref{fig:cone} for us. Y.~B.~G. is supported by the Adams Fellowship Programme of the Israeli Academy of Sciences and Humanities. V.~D. and Y.~B.~G. acknowledge support from the Israel Science Foundation (grant no. 2562/20). D.~J. acknowledges support from NASA ATP program (80NSSC18K1103). F.~S. acknowledges support from the Starting Grant (ERC-2015-STG 678652) ``GrInflaGal'' of the European Research Council.

\appendix

\section{Alternative derivation of the inferred number counts}
\label{sec:omega_tau in terms of inferred coordinates}

Here, we derive $i^*(\star j)$ from the volume form $\omtau$ expressed with the null tetrad (equation \eqref{eq:invariant light cone volume form}) in order to exemplify the role of the light-cone structure. This derivation is entirely independent of the choice of coordinates on the past light cone $\Sigtau$.

For this purpose, let us define the Levi-Civita tensor by\footnote{See  \cite[vol. 1, chapter 3]{Penrose:1984uia}) and \cite[chapter 8]{misner/thorne/wheeler} for technicalities.}
\begin{align}
    &e_{\mu\nu\rho\lambda} = \sqrt{\abs{\det g}}\eps_{\mu\nu\rho\lambda}\;,\qquad
    e^{\mu\nu\rho\lambda} = -\frac{1}{\sqrt{\abs{\det g}}}\eps^{\mu\nu\rho\lambda}
\end{align}
where $\varepsilon^{\mu\nu\rho\lambda}$ is the usual permutation symbol.
Furthermore, let
\begin{equation}
    A_{[\mu_1\dots\mu_n]} = \frac{1}{n!} \varepsilon_{\mu_1\dots\mu_n} \sum_\text{perms} A_{\mu_1\dots\mu_n}
\end{equation}
denote the anti-symmetric part of a tensor $A_{\mu_1\dots \mu_n}$.
Since the tensor $k_{[\mu}l_\nu m_\rho\bar{m}_{\lambda]}$ is a rank-4 fully anti-symmetric tensor (which follows from the properties of the null tetrad), this implies the existence of a scalar function $f(x)$ such that $k_{[\mu}l_\nu m_\rho\bar{m}_{\lambda]} = fe_{\mu\nu\rho\lambda}$. Multiplying both sides by $e^{\mu\nu\rho\lambda}$ yields
\begin{equation}
    -24f = f e_{\mu\nu\rho\lambda}e^{\mu\nu\rho\lambda} = k_\mu l_\nu m_\rho\bar{m}_\lambda e^{\mu\nu\rho\lambda} = \mathrm{i}\;,
\end{equation}
whence $f = -\frac{\mathrm{i}}{24}$.
Consider now the restriction of $\mathrm{i}(k\cdot u_g)l_\nu m_\rho \bar m_\lambda\mathrm{d}x^\nu\wedge\mathrm{d}x^\rho\wedge\mathrm{d}x^\lambda$
to the hyper-surface $\Sigtau$ parameterised with the inferred coordinates $\tilde x^i$ or, equivalently, the pull-back of $(k\cdot u_g) \omtau$ to $\Sigtau$.
Exploiting the commutativity of the pull-back $i^*$ with exterior products and differentiation, we have the following equalities:
\begin{equation}
\begin{aligned}
i^*\big[\big(k\cdot u_g\big)\omtau\big] &= i^*\Big[\big(k\cdot u_g\big)
(-\mathrm{i})\big(-l_\nu \mathrm{d}x^\nu\big)\wedge\big(\bar m_\rho \mathrm{d}x^\rho\big)\wedge\big(m_\lambda \mathrm{d}x^\lambda\big)\Big] \\
&= \mathrm{i}\, k_\mu u_g^\mu l_\nu m_\rho\bar{m}_\lambda~ i^*\!\big(\mathrm{d}x^\nu\big)\wedge i^*\!\big(\mathrm{d}x^\rho\big)\wedge i^*\!\big(\mathrm{d}x^\lambda\big) \Big. \\
&= \mathrm{i}\big(u_g^\mu k_\mu\big)\,l_{[\nu}m_\rho\bar{m}_{\lambda]}\, i^*\!\big(\mathrm{d}x^\nu\big)\wedge i^*\!\big(\mathrm{d}x^\rho\big)\wedge i^*\!\big(\mathrm{d}x^\lambda\big) \Big. \\ &
    = \mathrm{i}u_g^\mu\Big(k_\mu l_{[\nu}m_\rho\bar{m}_{\lambda]} + k_\nu l_{[\rho} m_\lambda \bar m_{\mu]} + \mbox{2 cyc.} \Big)\,i^*\!\big(\mathrm{d}x^\nu\big)\wedge i^*\!\big(\mathrm{d}x^\rho\big)\wedge i^*\!\big(\mathrm{d}x^\lambda\big) \Big. \\
    &= 4\mathrm{i}u_g^\mu\, k_{[\mu}l_\nu m_\rho\bar{m}_{\lambda]}\, i^*\!\big(\mathrm{d}x^\nu\big)\wedge i^*\!\big(\mathrm{d}x^\rho\big)\wedge i^*\!\big(\mathrm{d}x^\lambda\big) \Big. \\
    &= \bigg. \frac{1}{3!} u_g^\mu\, e_{\mu\nu\rho\lambda}~ i^*\!\big(\mathrm{d}x^\nu\big)\wedge i^*\!\big(\mathrm{d}x^\rho\big)\wedge i^*\!\big(\mathrm{d}x^\lambda\big) \\
    &= \frac{1}{3!}\sqrt{|\det g|} \,\varepsilon_{\mu\nu\rho\lambda}\, u_g^\mu\, i^*\!\big(\mathrm{d}x^\nu\big)\wedge i^*\!\big(\mathrm{d}x^\rho\big)\wedge i^*\!\big(\mathrm{d}x^\lambda\big) \;.
\end{aligned}
\end{equation}
In the fourth equality, we took advantage of
\begin{equation}
k_\mu\, i^*\big(\mathrm{d}x^\mu\big) = i^*\big(k_\mu \mathrm{d}x^\mu\big)
= -i^*\big(\mathrm{d}V\big) = 0 \;,
\end{equation}
which implies that all the terms added on the fourth line cancel out for they involve $k_\mu \mathrm{d}x^\mu \equiv - \mathrm{d}V$ that vanishes on $\Sigtau$ by definition. This stresses that the pull-back $i^*$ plays a crucial role in this derivation.

Finally, on substituting the inferred coordinates $\tilde x^i$, the pull-back of $(k\cdot u_g)\omtau$ becomes
\begin{equation}
\label{eq:k.u omega tau app}
  i^*\big[\big(k\cdot u_g\big)\omtau\big] = \sqrt{|\det g|}\, \varepsilon_{\mu\nu\rho\lambda}\, u_g^\mu(\tvx)\,
  \frac{\partial x^\nu}{\partial \tilde{x}^1}\frac{\partial x^\rho}{\partial \tilde{x}^2}\frac{\partial x^\lambda}{\partial \tilde{x}^3}
  \,\mathrm{d}\tilde x^1\wedge\mathrm{d}\tilde x^2\wedge\mathrm{d}\tilde x^3
  \;,
\end{equation}
which is \refeq{k.u omega tau}, and corresponds to the well-known expression for the (physical) volume measured by an observer with 4-velocity $u_g^\mu$ \citep{weinberg:1972}.
We recover again equation Eq.~\eqref{eq: star j resticted} upon multiplication by $n_g(\tvx)$.

\section{Alternative derivation of the observed, linear galaxy over-density}
\label{sec:alternative}

Starting from the perturbed FLRW metric \eqref{eq:perturbed metric}, the ingredients of the galaxy number counts expressed in the form (\ref{eq:galaxy number counts coordinates}) are given by
\begin{equation}
\begin{aligned}
  & \sqrt{\abs{\det g}} = \left(1+A+\frac{h}{2}\right)a^3 \\ &
  \eps_{abcd}u^a\frac{\partial x^b}{\partial \tilde{x}^1}\frac{\partial x^c}{\partial \tilde{x}^2}\frac{\partial x^d}{\partial \tilde{x}^3} = \big(1-A\big)\big(1+\tr D\big) + v_\parallel
\end{aligned}
\end{equation}
to first order in perturbations.
Here, $h = \tr (h_{ij})$, $D_{ij} = \frac{\partial x^i}{\partial \tilde{x}^j} - \delta^i_j$, and we used the approximation $\det (I+M) \approx 1 + \tr M$ valid to linear order in any matrix $M$.
Equating equations \eqref{eq:galaxy number counts coordinates} and \eqref{eq:number counts in inferred coordinates} and substituting the expressions we found for the determinants, we arrive at
\begin{equation}\label{eq: over-denisty intermediate}
  \frac{\tilde{n}_g(\tvx)}{n_g(\tvx)} = \left(1+A+\frac{h}{2}\right)\frac{a^3}{\tilde{a}^3}\left(1-A + \tr D + v_\parallel\right) \;.
\end{equation}
What remains to be done is to compute $\tr D$ and to relate $n_g(\tilde x)$ to the average galaxy density at fixed observed redshift. At this point, it is important to remember that $V=$~const hyper-surfaces are not parallel with constant conformal time, $\eta=$~const, hyper-surfaces. This implies that $\frac{\partial}{\partial \tilde{x}^i}$ is not parallel to $\frac{\partial}{\partial x^i}$. We thus write $\tilde{\mathbf{x}} = \mathbf{x} + \Delta \mathbf{x}$, and $\Delta \mathbf{x} = \Delta x_\parallel \nvh + \Delta \mathbf{x}_\perp$, with $\nvh \cdot \Delta \mathbf{x}_\perp = 0$ and $\ntvh\equiv\nvh$ (the sky direction seen by the observer in linear theory).
In the notation of Ref.~\cite{SchmidtJeong2012}, if $P^{ij} = \delta^{ij} - \nh^i\nh^j$ is the projection tensor, then $\Delta x_\perp^i = P^{i}_j\Delta x^j$. Let us proceed to calculate the Hessian $D$:
\begin{equation}
  \frac{\partial \tilde{x}^i}{\partial x^j} = \delta^i_j -\frac{\partial \Delta x^i}{\partial x^j}\;,
\end{equation}
whence
\begin{equation}
\begin{aligned}
  \tr D & = \sum_{i=1}^{3}\frac{\partial \Delta x^i}{\partial x^i} = \frac{\partial(\nh_i\Delta x^i)}{\partial x^i} + \frac{\partial \Delta x_\perp^i}{\partial x^i} \\ &
  = \partial_\parallel \Delta x_\parallel + \Delta x_\parallel \frac{\partial \nh^i}{\partial x^i}  + \frac{\partial(P^{ik}P_k^j\Delta x_j)}{\partial x^i} \\ &
  = \partial_\parallel\Delta x_\parallel + \frac{2}{\chi}\Delta x_\parallel + \Delta x_{\perp \;k}\frac{\partial P^{ik}}{\partial x^i} + P^{ik}\frac{\partial \Delta x_{\perp \; k}}{\partial x^i} \\ &
  = \partial_\parallel\Delta x_\parallel + \frac{2}{\chi}\Delta x_\parallel + \partial_{\perp\;k}\Delta x_{\perp \; k} \\ &
  = \partial_\parallel\Delta x_\parallel + \frac{2}{\chi}\Delta x_\parallel -2\hat{\kappa}\;.
\end{aligned}
\end{equation}
In the above expressions, we used the relations $\frac{\partial\nh^i}{\partial x^i} = \frac{2}{\chi}$, and $\frac{\partial }{\partial x^j}P^{ki} = -\frac{1}{\chi}\left(P_j^k\nh^i + P_j^i\nh^k\right)$. We have also defined
\begin{equation}
  \hat{\kappa} = -\frac{1}{2}\partial_{\perp\;k}\Delta x_{\perp \; k} = -v_{\parallel o}+\kappa\;,
\end{equation}
where $\kappa$ is the lensing convergence, Eq.~\eqref{eq:kappa}.
At zeroth order, one has $\partial_\parallel = \partial_{\tilde{\chi}}$, since the two are derivatives along the line of sight. Moreover, as $\tr D$ is already first order, one may replace $\partial_\parallel$ with $\partial_{\tilde{\chi}}$ in that expression to obtain
\begin{equation}
  \tr D = \partial_{\tilde{\chi}}\Delta x_\parallel + \frac{2}{\tilde{\chi}}\Delta x_\parallel -2\hat{\kappa}\;.
\end{equation}
As for the physical galaxy density $n_g(\tvx)$ (pulled back onto the past light cone $\Sigtau$), we follow Ref.~\cite{JeongSchmidt2015} and choose the constant observed redshift gauge. The requirement $n_g(\tvx) \equiv \bar{n}_g(\tilde{z})(1+\delta^{\textrm{or}}_g)$ defines the fractional galaxy perturbation $\delta_g^\text{or}$ at constant $\tilde z$. With the choice of local coordinates $\tilde x^i=(\tilde z,\ntvh)$ for instance, Eq.~\eqref{eq: over-denisty intermediate} reads
\begin{equation}
  \frac{\tilde{n}_g(\tilde{z},\ntvh)}{\bar{n}(\tilde{z})} = \left(1+A+\frac{h}{2}\right)\frac{a^3}{\tilde{a}^3}\left(1-A + \partial_{\tilde{\chi}}\Delta x_\parallel + \frac{2}{\tilde{\chi}}\Delta x_\parallel -2\hat{\kappa} + v_\parallel\right)\left(1+\delta^{\textrm{or}}_g\right).
\end{equation}
Substituting $\Delta\!\ln\! a=\frac{a}{\tilde{a}}-1$ and linearising the previous expression yields
\begin{equation}
   \frac{\tilde{n}_g(\tilde{z},\ntvh)}{\bar{n}(\tilde{z})} = 1+\frac{h}{2} + 3\, \Delta\!\ln\! a + \partial_{\tilde{\chi}}\Delta x_\parallel + \frac{2}{\tilde{\chi}}\Delta x_\parallel -2\hat{\kappa} + v_\parallel +\delta^{\textrm{or}}_g.
\end{equation}
We can now express $\dgobs(\tilde z,\ntvh)= \frac{\tilde{n}_g(\tilde{z},\ntvh)}{\bar{n}(\tilde{z})} - 1$ in terms of gauge-invariant quantities alone. On inserting the definition of the gauge-invariant rulers $\mathcal{C}$, $\mathcal{M}$, and $\mathcal{T}$, we eventually obtain Eq.~\eqref{eq:over density linear}.

\section{Gauge-invariance of $\mathcal{C}$}
\label{sec:C gauge invariance}

Consider the linear expression of $\mathcal{C}$ for the perturbed FLRW metric in a general gauge \eqref{eq:perturbed metric}:
\begin{equation}
\begin{aligned}
    \mathcal{C} & = \frac{\mathrm{d}\ln r_0(\tilde{a})}{\mathrm{d}\!\ln\!\tilde{a}} \mathcal{T} + \left(\frac{\dot{\tilde{\cal H}}}{\tilde{\cal H}^2}- 1\right)\Delta\!\ln\! a - A - v_\parallel + B_\parallel \\ &
    \qquad + \frac{1}{\tilde{\cal H}}\left(-\partial_\parallel A + \partial_\parallel v_\parallel + \dot{B}_\parallel - \dot{v}_\parallel +\frac{1}{2}\dot{h}_\parallel\right) \;,
\end{aligned}
\end{equation}
where $\mathcal{T}$ is given by equation (15) of \cite{JeongSchmidt2014}, a tilde denotes background quantities evaluated at the observed redshift $\tilde z$, a dot denotes a derivative w.r.t. the conformal time, and
\begin{equation}
    \Delta\! \ln\! a = A_o - A + v_\parallel - v_{\parallel o} + \int_0^{\tilde{\chi}}\mathrm{d}\chi ~\left(-\dot{A} + \frac{1}{2}\dot{h}_\parallel + \dot{B}_\parallel\right) - H_0\int_0^{\bar\eta_0}\mathrm{d}\eta~A(\boldsymbol{0},\eta)a(\eta) \;.
\end{equation}
All the source and observer terms are evaluated at the spacetime coordinates $(\tilde\eta,\tilde\chi\ntvh)$ and $(\bar\eta_o,{\bf 0})$, respectively.
The quantity $\frac{\mathrm{d} \ln r_0(\tilde{a})}{\mathrm{d} \ln \tilde{a}}$ is the intrinsic change in the length of the ruler due to a change in scale factor (corresponding to the observed redshift $\tilde z$) which is discussed in Section \ref{sec:non-linearT}, and is gauge-invariant \emph{per definitionem}. The $\mathcal{T}$-ruler is also gauge-invariant as demonstrated in \cite{JeongSchmidt2014}.

Here, we show explicitly that $\mathcal{C}$ is invariant under a (scalar) gauge transformation
\begin{equation}
    x^a \mapsto x^a + \left(\begin{array}{c}
        T \\
        \partial^{i}L
    \end{array}\right).
\end{equation}
The corresponding transformations of the metric elements are specified in equation (A2) of \cite{JeongSchmidt2014}. They imply $\Delta\! \ln\! a \mapsto \Delta \!\ln\! a + \tilde{\cal{H}}\tilde T$, where $\tilde T\equiv T(\tilde\eta,\tilde\chi\nvh)$. Therefore, $\mathcal C$ transforms as
\begin{align}
    \mathcal{C} &\mapsto  \mathcal{C} + \left(\frac{\dot{\tilde{\cal H}}}{\tilde{\cal H}^2}- 1\right)\tilde{\cal H}\tilde{T}+ \tilde{\cal H}\tilde T + \dot{\tilde T} - \partial_\parallel \dot{\tilde L}
    + \partial_\parallel\dot{\tilde L} - \partial_\parallel\tilde T \nonumber \\ &
    \quad + \frac{1}{\tilde{\mathcal{H}}}\Big[\mathcal{H}\,\partial_\parallel\tilde T + \partial_\parallel \dot{\tilde T} + \partial^2_\parallel \dot{\tilde L} + \partial_\parallel \ddot{\tilde L} - \partial_\parallel \ddot{\tilde L} - \partial_\parallel\tilde T - ({\dot{\tilde{\cal H}}} \tilde T+\tilde{\cal H}\dot{\tilde T}) - \partial_\parallel^2 \dot{\tilde L}\Big] \nonumber \\ &
    \quad = \mathcal{C} + \frac{\dot{\tilde{\cal H}}}{\tilde{\cal H}} T + \dot{\tilde T} -\partial_\parallel\tilde T + \frac{1}{\tilde{\mathcal{H}}}\Big(\tilde{\cal H}\,\partial_\parallel\tilde T - {\dot{\tilde{\cal H}}} T - \tilde{\cal H}\dot{\tilde T}\Big) \nonumber \\ & \quad = \mathcal{C} \;,
\end{align}
where $\tilde L\equiv L(\tilde\eta,\tilde\chi\nvh)$ and $\partial_\parallel = \nh^i\partial_i$.

\section{Angular Power Spectrum in Einstein-de-Sitter Background}
\label{sec:rulers in EdS}

In an Einstein-de Sitter (EdS) background, the Bardeen potentials satisfy $\Phi = \Psi$ and both are independent of time. Using the fact that $\frac{\dot{\tilde{\cal H}}}{\tilde{\cal H}^2}-1 = -\frac{3}{2}$, the parallel and transverse ruler quantities given by Eqs.~\eqref{eq:CrulercN} and \eqref{eq:MrulercN} simplify to
\begin{align}
   \left(\mathcal{C} - \frac{\mathrm{d}\ln r_0}{\mathrm{d}\ln \tilde{a}}\mathcal{T}\right)_\text{cN} & = \frac{1}{2}\big(\Psi - \Psi_o\big) - \frac{5}{2}\big(v_\parallel - v_{\parallel,o}\big) + \frac{1}{\mathcal{H}}\big(-\partial_\parallel\Psi + \partial_\parallel v_\parallel - \dot{v}_\parallel\big) - v_{\parallel o} \nonumber \\
\left(\mathcal{M} - 2\frac{\mathrm{d}\ln r_0}{\mathrm{d}\ln \tilde{a}}\mathcal{T}\right)_\text{cN} &=
    \left(-2+\frac{2}{\tilde{\cal H}\tilde\chi}\right)\big(v_\parallel-v_{\parallel o}+\Psi_o-\Psi\big) + 2 \big(\Psi-\Psi_o\big) \nonumber \\ &\qquad + 2\big(-v_{\parallel o} + \kappa\big) -\frac{4}{\tilde\chi}\int_0^{\tilde\chi}\!\mathrm{d}\chi\,\big(\Psi-\Psi_o\big)\;,
\end{align}
where
\begin{equation}
  \kappa = \frac{1}{\tilde\chi} \int_{0}^{\tilde{\chi}}\!\mathrm{d}\chi\,\chi(\tilde{\chi}-\chi)\partial_\perp^2\Psi \;.
\end{equation}
Note the cancellation that occurs between $-2\left(\frac{\dot{\cal H}_o}{{\cal H}_o}-{\cal H}_o\right)\delta\eta_o$ and $2\Psi_o$ , so that $\mathcal{H}_o$ and $\HF$ only differ by $+\frac{1}{3\mathcal{H}_o}\partial_i v_o^i$ in EdS.
Furthermore, the expression \eqref{eq:TrulercN} of the $\mathcal T$-ruler reduces to
\begin{equation}
  \big(\mathcal{T}\big)_\text{cN} = \frac{1}{3}\big(\Psi_o - \Psi\big) + v_\parallel - v_{\parallel,o} \;.
\end{equation}
These expressions recover those presented in \cite{SchmidtJeong2012,JeongSchmidt2014}, which emphasises that the two possible conventions for the $\mathcal{C}$ and $\mathcal{M}$ rulers presented in Section \ref{sec:delta obs with cN rulers} agree in this limit.
They exemplify the property that the rulers are unaffected either by a constant or by a pure gradient potential shift.\footnote{In a generic cosmology, the invariance of the rulers under the addition of a long mode (that is, $\Psi\to \Psi+\Psi_\ell$ and $\Phi\to\Phi+\Phi_\ell$) can be demonstrated along the lines of \cite{Peloso:2013zw,Kehagias:2013yd,Creminelli:2013mca}}

On taking the Fourier transform and using Rayleigh's identity, the various components of the above expressions can be decomposed in a spherical Fourier basis, \emph{viz.}
\begin{align}
  \big(v_\parallel\big)_{lm}\!(\tilde{z}) & = 4\pi \dot{\tilde{a}}\,\mathrm{i}^l\intk j_l'(k\tilde{\chi})Y_{lm}^*\!\big(\kvh\big)\frac{\delta_m(\bar\eta_o,\vk)}{k} \\
  \big(\Psi\big)_{lm}\!(\tilde{z}) & = 6\pi \mathcal{H}_o^2\,\mathrm{i}^l\intk j_l(k\tilde{\chi})Y_{lm}^*\!\big(\kvh\big)\frac{\delta_m(\bar\eta_o,\vk)}{k^2} \nonumber \\
  \big(\partial_\parallel v_\parallel\big)_{lm}\!(\tilde{z}) & = 4\pi \dot{\tilde{a}}\,\mathrm{i}^l\intk j_l''(k\tilde{\chi})Y_{lm}^*\!\big(\kvh\big)\delta_m(\bar\eta_o,\vk) \nonumber \\
  \big(\dot{v}_\parallel\big)_{lm}\!(\tilde{z}) & = 4\pi \ddot{\tilde{a}}\,\mathrm{i}^l\intk j_l'(k\tilde{\chi})Y_{lm}^*\!\big(\kvh\big)\frac{\delta_m(\bar\eta_o,\vk)}{k} \nonumber \\
  \left(\frac{4}{\tilde{\chi}}\int_{0}^{\tilde{\chi}}\!\mathrm{d}\chi\,\Psi\right)_{lm}\!(\tilde{z}) & = 24\pi \mathcal{H}_o^2\,\mathrm{i}^l\intk \frac{\int_0^{k\tilde{\chi}} j_l(x)\mathrm{d}x}{k\tilde{\chi}}Y_{lm}^*\!\big(\kvh\big)\frac{\delta_m(\bar\eta_o,\vk)}{k^2} \nonumber \\
  \big(\kappa\big)_{lm}\!(\tilde{z}) & = 6\pi \mathcal{H}_o^2\,\mathrm{i}^l\intk \,
  \left[\int_{0}^{\tilde{\chi}} \mathrm{d}\chi \frac{\chi(\tilde{\chi}-\chi)}{\tilde{\chi}}\left(j_l''(k\chi) - j_l(k\chi) + \frac{2}{k\chi}j_l'(k\chi)\right)\right] \nonumber \\ & \qquad \times
  Y_{lm}^*\!\big(\kvh\big)\delta_m(\bar\eta_o,\vk) \nonumber \;,
\end{align}
where $\intk\equiv(2\pi)^{-3}\int\!\mathrm{d}^3k$, a prime denotes a derivative with respect to the argument of the function, and the present-day, Fourier mode $\delta_m(\bar\eta_o,\vk)$ of matter fluctuations is in synchronous comoving gauge. Substituting all the terms above into the cN expressions of $\mathcal{C}$ and $\mathcal{M}$, we can express the multipoles of the observed galaxy over-density as
\begin{equation}
    \big(\dgobs\big)_{lm}(\tilde z) = 4\pi \mathrm{i}^l\int_{\vk} F_l(k\tilde{\chi})\,Y_{lm}^*\big(\kvh\big)\,\delta_m(\bar\eta_o,\vk)\;,
\end{equation}
where the real-valued function $F_l(k\tilde\chi)$ is a suitable superposition of Bessel functions, \emph{etc.} Note that the decomposition can also be carried out with spherical Fourier Bessel functions \cite[see, e.g.,][]{yoo/desjacques:2013}, in which case the dependence is on the inferred wave number $\tilde k$ instead of $\tilde z$.


\section{List of Symbols}
\label{appendix:symbol list}

\begin{center}
\begin{longtable}{|c|l|c|}
\hline
  Symbol & Meaning & Equation\\
  \hline
  $\tilde{a}$ & $1/(1+\tilde{z})$ & below eq. \eqref{eqn: inferred 3-volume form} \\
  $\at(\tau)$ & scale factor evaluated at source proper time $\tau$ & above eq. \eqref{eq:defTa} \\
  $\mathcal{A}_{ij}$ & transverse cosmic ruler & \eqref{eq:ruler definitions} \\
  $b_1$ & linear galaxy bias & \eqref{eqn: equation 4.12} \\
  $\mathcal{B}_i$ & vector cosmic ruler & \eqref{eq:ruler definitions} \\
  $\mathcal{C}$ & longitudinal cosmic ruler & \eqref{eq:ruler definitions} \\
  $\tilde\chi$ & $\chi(\tilde z)$, the inferred comoving distance to redshift $\tilde z$ & below eq. \eqref{eq:redshift} \\
  $\delta\eta$ & conformal time perturbation & above eq. \eqref{eq:as} \\
  $\dgobs$ & observed galaxy over-density & \eqref{eq:def of observed delta_g} \\
  $\delta_g^\textrm{or}$ & galaxy over-density on constant redshift slices & \eqref{eq:over density non linear} \\
  $\delta_g^\textrm{pt}$ & galaxy over-density on constant proper-time slices & \eqref{eq:or-to-pt} \\
  $\delta_m^\textrm{syn}$ & matter over-density on constant coordinate time slice & \eqref{eqn: equation 4.12} \\
  $\det\!\mathcal D_o$ & Jacobi map determinant & \eqref{eq:detjacobimap} \\
  $\eta$ & conformal time coordinate & \eqref{eq:perturbed metric} \\
  $\bar\eta(\tilde z)$ & background redshift-to-conformal-time relation & above eq. \eqref{eq:as} \\
  $\overline{g}(\tilde{a})$ & FLRW background evaluated at $\tilde{a}$ & below eq. \eqref{eqn: inferred 3-volume form} \\
  $\tilde{H}$ & background expansion rate evaluated at $\tilde a$ & \eqref{eqn: equation 3.25} \\
  $\tilde{\mathcal{H}}$ & background conformal Hubble rate evaluated at $\tilde a$ & \eqref{eqn: equation 3.25} \\
  $\HF$ & CFC Hubble rate & \eqref{eq:HF} \\
  $j$ & galaxy current 4-vector & \eqref{eq:galaxy number count Hodge} \\
  $\mathcal{J}_{\mathcal{T}}$ & Cosmic clock Jacobian & \eqref{eq:JT} \\
  $k$ & null tetrad element, light wave-vector & \eqref{eq:nulltetrad} \\
  $l$ & null tetrad element, perpendicular to $\Sigtau$ & \eqref{eq:nulltetrad} \\
  $\lambda$ & light-ray affine parameter & \eqref{eq: proper length along light ray} \\
  $\mathcal{M}$ & $\tr \mathcal{A}_{ij}$ & \eqref{eq:over density linear} \\
  $m$ & transverse null tetrad element & \eqref{eq:nulltetrad} \\
  $\bar{m}$ & transverse null tetrad element & \eqref{eq:nulltetrad} \\
  \hline
  $n_g$ & physical number density of galaxies in their rest frame & above eq. \eqref{eq:galaxy number count Hodge} \\
  $n_{\rm g,lc}$ & invariant number-density on $\Sigtau$, equal to $n_g k \cdot u_g$ & \eqref{eq:galaxy number count} \\
  $\tilde n_g(\tvx)$ & inferred galaxy density & \eqref{eq:def of observed delta_g} \\
  $\bar n_g(\tilde z)$ & mean physical galaxy number density at observed redshift $\tilde z$ & \eqref{eq:def of observed delta_g} \\
  $\ntvh$ & observed sky direction & below eq. \eqref{eqn: inferred 3-volume form} \\
  $\omega$ & volume form space-time & \eqref{eq:full volume form} \\
  $\omtau$ & volume form of $\Sigtau$ & \eqref{eq:galaxy number count} \\
  $\tomtau$ & inferred 3-volume form & \eqref{eqn: inferred 3-volume form} \\
  $\Psi$ & `temporal' Bardeen potential & below eq. \eqref{eq:perturbed metric} \\
  $\Phi$ & `spatial' Bardeen potential & below eq. \eqref{eq:perturbed metric} \\
  $\bm{r}_{0}$ & intrinsic ruler size & \eqref{eq:ruler definitions} \\
  $r_0(\tau)$ & intrinsic ruler size as function of its proper time & \eqref{eqn: equation 3.1} \\
  $\tilde{\bm{r}}$ & inferred ruler size & \eqref{eq:ruler definitions} \\
  $\sigma$ & invariant, transverse area $2$-form & \eqref{eq:galaxy number counts forms final} \\
  $\Sigma$ & Null hyper-surface & \eqref{eq:galaxy number count Hodge} \\
  $\Sigtau$ & Observer's past light-cone & \eqref{eq:galaxy number count} \\
  $\mathcal{T}_a$ & cosmic clock & \eqref{eq:defTa} \\
  $\tau_s$ & proper time of the source, as measured by it & \eqref{eq:JT} \\
  $\overline{\tau}(\tilde a)$ & background proper time-redshift relation & \eqref{eq:JT} \\
  $u_g$ & galaxy 4-velocity & \eqref{eq:invariant number density} \\
  $\mathrm{d}U$ & dual to $k$ & \eqref{eq:full volume form} \\
  $V_0(\tau_s)$ & ruler volume as function of source proper time & \eqref{eq:V0_Ta} \\
  $V_0(\tilde a)$ & shorthand for $V_0(\overline\tau(\tilde a))$ & \eqref{eq:V0_Ta} \\
  $\mathrm{d}V$ & dual to $l$ & \eqref{eq:full volume form} \\
  $\tilde{z}$ & observed redshift & \eqref{eqn: dN by dz 2.15} \\
  $\mathrm{d}\zeta$ & dual to $m$ & \eqref{eq:full volume form} \\
  $\mathrm{d}\bar{\zeta}$ & dual to $\bar{m}$ & \eqref{eq:full volume form} \\
  \hline
\end{longtable}
\end{center}

\bibliographystyle{JHEP}
\bibliography{references}

\label{lastpage}

\end{document}